\documentclass[a4paper,11pt]{article}

\usepackage{geometry}										
\usepackage[english]{babel}									
\usepackage[utf8]{inputenc}									
\usepackage[T1]{fontenc}									

\usepackage{lmodern}										

\usepackage{amsmath}										
\usepackage{amstext}										
\usepackage{amsfonts}										
\usepackage{amssymb}										

\usepackage{bm}											
\usepackage{dsfont}										
\usepackage{units}										

\usepackage{textcomp}										

\usepackage{graphicx}										
\usepackage[svgnames]{xcolor}									
\usepackage{booktabs}										
\usepackage{multirow}										
\usepackage[inline]{enumitem}
\usepackage{algorithm}
\usepackage{algorithmic}
\usepackage{appendix}

\graphicspath{{./fig/}}

\usepackage{mathtools}
\newcommand{\caa}{{\mathcal A}}
\newcommand{\cd}{{\mathcal D}}
\newcommand{\ch}{{\mathcal H}}
\newcommand{\cl}{{\mathcal L}}
\newcommand{\cm}{{\mathcal M}}
\newcommand{\cn}{{\mathcal N}}
\newcommand{\cu}{{\mathcal U}}
\newcommand{\cx}{{\mathcal X}}
\newcommand{\cy}{{\mathcal Y}}
\newcommand{\Rr}{{\mathbb R}}
\newcommand{\Nn}{{\mathbb N}}

\newcommand{\mathbbm}[1]{\text{\usefont{U}{bbm}{m}{n}#1}} 
\newcommand{\Esp}[1]{{\mathbb E}\left[ #1 \right]}
\newcommand{\Espe}[2]{{\mathbb E}_{#1}\left[#2\right]}
\newcommand{\Var}[1]{{\rm Var}\left[ #1 \right]}
\newcommand{\Cov}[1]{{\rm Cov}\left[ #1 \right]}
\newcommand{\ve}[1]{\boldsymbol{#1}}
\newcommand{\acc}[1]{\left\{#1\right\}}
\newcommand{\eqdef}{\stackrel{\text{def}}{=}}

\newcommand{\D}{\mathrm{d}}
\newcommand{\support}[1]{\suppt\left(#1\right)}

\DeclareMathOperator*{\PC}{PC}
\DeclareMathOperator{\suppt}{supp}
\DeclareMathOperator{\GLD}{GLD}

\DeclareMathOperator{\Betafun}{B}
\DeclareMathOperator{\LAR}{Hybrid-LAR}
\DeclarePairedDelimiter\norm{\lVert}{\rVert}

\newcommand{\refEq}[1]{Eq.~(\ref{#1})}
\newcommand{\refEqs}[2]{Eqs.~(\ref{#1})-(\ref{#2})}

\newcommand{\bestresult}[1]{\textbf{#1}}
\newcommand{\GMM}{\textit{GLD~MM}}
\newcommand{\GMLE}{\textit{GLD~MLE}}
\newcommand{\GJMM}{\textit{GLD~joint\_MM}}
\newcommand{\GJMLE}{\textit{GLD~joint\_MLE}}
\providecommand{\ie}{i.e.,\;}
\providecommand{\eg}{e.g.,\;}

\newlength{\HYDROsubWidth}	
\newlength{\HYDROfigHeight}	
\newlength{\HYDROmapHeight}	
\newlength{\HYDROfigHeightNew}
\usepackage{authblk}										

\title{Replication-based emulation of the response distribution of stochastic simulators using generalized lambda distributions}
\author[1]{Xujia Zhu \thanks{zhu@ibk.baug.ethz.ch}}
\author[1]{Bruno Sudret\thanks{sudret@ethz.ch}}
\affil[1]{Chair of Risk, Safety and Uncertainty Quantification, ETH Z\"{u}rich, Stefano-Franscini-Platz 5, 8093 Z\"{u}rich, Switzerland}

\date{\today}

\usepackage{microtype}										

\usepackage{indentfirst}									

\usepackage{caption}										
\usepackage{subfigure}

\usepackage[numbers,sort&compress]{natbib}							
\usepackage{hyperref}
\hypersetup{
pdftitle = {Replication-based emulation of the response distribution of stochastic simulators using generalized lambda distributions},
pdfauthor = {Xujia Zhu, Bruno Sudret},
pdfsubject = {Manuscript submitted to Int. J. Uncertain. Quantificat.},
pdfkeywords = {Stochastic simulators, Surrogate modeling, Generalized lambda distributions, Sparse polynomial chaos expansions},
colorlinks = false,										
}

\usepackage[capitalize]{cleveref}								
\creflabelformat{equation}{#2(#1)#3}								
\crefrangelabelformat{equation}{#3(#1)#4 to #5(#2)#6}						
\labelcrefformat{subequation}{#2(#1)#3}								
\labelcrefrangeformat{subequation}{#3(#1)#4 to #5(#2)#6}					

\begin{document}

\maketitle

\begin{abstract}
Due to limited computational power, performing uncertainty quantification analyses with complex computational models can be a challenging task. This is exacerbated in the context of stochastic simulators, the response of which to a given set of input parameters, rather than being a deterministic value, is a random variable with unknown probability density function (PDF). Of interest in this paper is the construction of a surrogate that can accurately predict this response PDF for any input parameters. We suggest using a flexible distribution family -- the generalized lambda distribution -- to approximate the response PDF. The associated distribution parameters are cast as functions of input parameters and represented by sparse polynomial chaos expansions. To build such a surrogate model, we propose an approach based on a local inference of the response PDF at each point of the experimental design based on replicated model evaluations. Two versions of this framework are proposed and compared on analytical examples and case studies.
\end{abstract}

\section{Introduction}
Computer models, a.k.a. simulators, are nowadays widely used in the context of design optimization, uncertainty quantification and sensitivity analysis. A simulator is called \emph{deterministic} if repeated runs with the same input parameters produce exactly the same output quantity of interest (QoI); for example, a finite element model of a structure with external load as input and stresses as output is a deterministic simulator. In contrast, a \emph{stochastic simulator} provides different results when run repeatedly with the same input values. In other words, for a given vector of input parameters, the QoI of a stochastic simulator is a \emph{random variable}, whose probability density function (PDF) is of interest. The reason for this intrinsic stochasticity is that some source of randomness inside the model, which can be represented by \emph{latent variables}, is not taken explicitly into account within the input parameters. Therefore, if not all the relevant variables that uniquely determine the output can be fully specified, the model output remains random. Examples of stochastic simulators are encountered when evaluating the performance of a wind turbine under stochastic loads when only some characteristic values of the wind climate are known, or when predicting the price of an option in financial market with only historical data.
\par
Such numerical models can be time-consuming: a single model evaluation may require minutes to hours of simulation, as it is the case for complex fluid dynamic codes. To alleviate the computational burden, surrogate models, a.k.a. emulators, have been successfully developed for deterministic simulators, such as Gaussian processes \citep{Rasmussen2006} and polynomial chaos expansions \citep{Xiu2002,Ghanembook2003}. The construction of surrogate models relies on a set of model evaluations, called the \emph{experimental design} (ED). However, when it comes to stochastic simulators, one single model evaluation for a given vector of input parameter is incapable to fully characterize the associated QoI. As a result, repeated runs with the same input parameters, called \emph{replications}, are necessary to obtain the resulting (unknown) probability distribution of the QoI. Consequently, standard surrogate modeling techniques cannot directly be applied to stochastic simulators, due to the very random nature of the output.
\par
Large efforts have been dedicated to estimate summary scalar quantities of the random output as a function of the input parameters, such as the mean value \citep{McCullagh1989,Iooss2009,Ankenman2009}, the standard deviation \citep{Dacidian1987,Fan1998I,Marrel2012} and quantiles \citep{Bhattacharya1990,Plumlee2014,Koenker2017}. However, surrogate modeling for the entire response PDF of a stochastic code is a less mature field. Two types of approaches can be found in the literature. The first is known as the \emph{statistical approach}. If the response PDF belongs to the exponential family, generalized linear models (GLM) can be efficiently applied \citep{McCullagh1989,Hastie1990}. When the probability distribution is arbitrary and no prior knowledge on its shape is available, nonparametric estimators may be considered, notably kernel density estimators \citep{Fan1996,Hall2004} and projection estimators \citep{Efromovich2010}. Nonparametric estimators, however, suffer from the \textit{curse of dimensionality} \citep{Tsybakov2009}, meaning that the necessary amount of data needed to achieve sufficient accuracy increases drastically with increasing input dimensionality. 
\par
A second approach is the \emph{replication-based} method, which capitalizes instead on available replications to represent the response distribution through a suitably general parametric distribution. The parameters of the latter are then treated as outputs of a deterministic simulator. Conventional deterministic surrogate modeling methods may then be used to emulate these parameters as functions of the input. Note that this approach was initially proposed to estimate summary statistics \citep{Ankenman2009,Plumlee2014}. It has been extended to more general cases given the functional form of the parametric PDF by \citep{Moutoussamy2015}. So far, nonparametric estimators have been used to estimate the distribution from replications \citep{Moutoussamy2015,Browne2016}. Thus, many replications are necessary, sometimes as many as $10^4$ replications per point of the experimental design, which severely limits the applicability of such an approach.
\par
It is worthwhile to notice that the existing methods either assume a rather restrictive shape of the distribution or require a large number of model evaluations. The present paper aims at designing a replication-based approach which will reduce the necessary amount of replications. To this end, we propose approximating the response PDF of a stochastic simulator by \emph{generalized lambda distributions} (GLD) \citep{Karian2000}. The distribution parameters are functions of the input and further represented by polynomial chaos (PC) expansions. To construct such a surrogate model, we present two algorithms in this paper: the first one follows the general idea of the replication-based approach, while the second enriches the former with an additional optimization step.
\par
The paper is organized as follows. \Cref{sec:GLD,sec:PCE} introduce generalized lambda distributions and polynomial chaos expansions, respectively. In \Cref{sec:JM}, we present our novel algorithms to infer the response PDF of a stochastic simulator based on limited replicated data. \Cref{sec:examples} validates the proposed methods through two toy examples, and \Cref{sec:appli} illustrates their performance on two applications, namely a stochastic differential equation case study and a wind turbine simulation. Finally, we summarize the main findings of the paper and provide outlooks for future research in \Cref{sec:conclusions}.

\section{Generalized lambda distributions}
\label{sec:GLD}

\subsection{Formulation}
\begin{figure}[htbp!]
	\centering
	\includegraphics[width=0.9\linewidth]{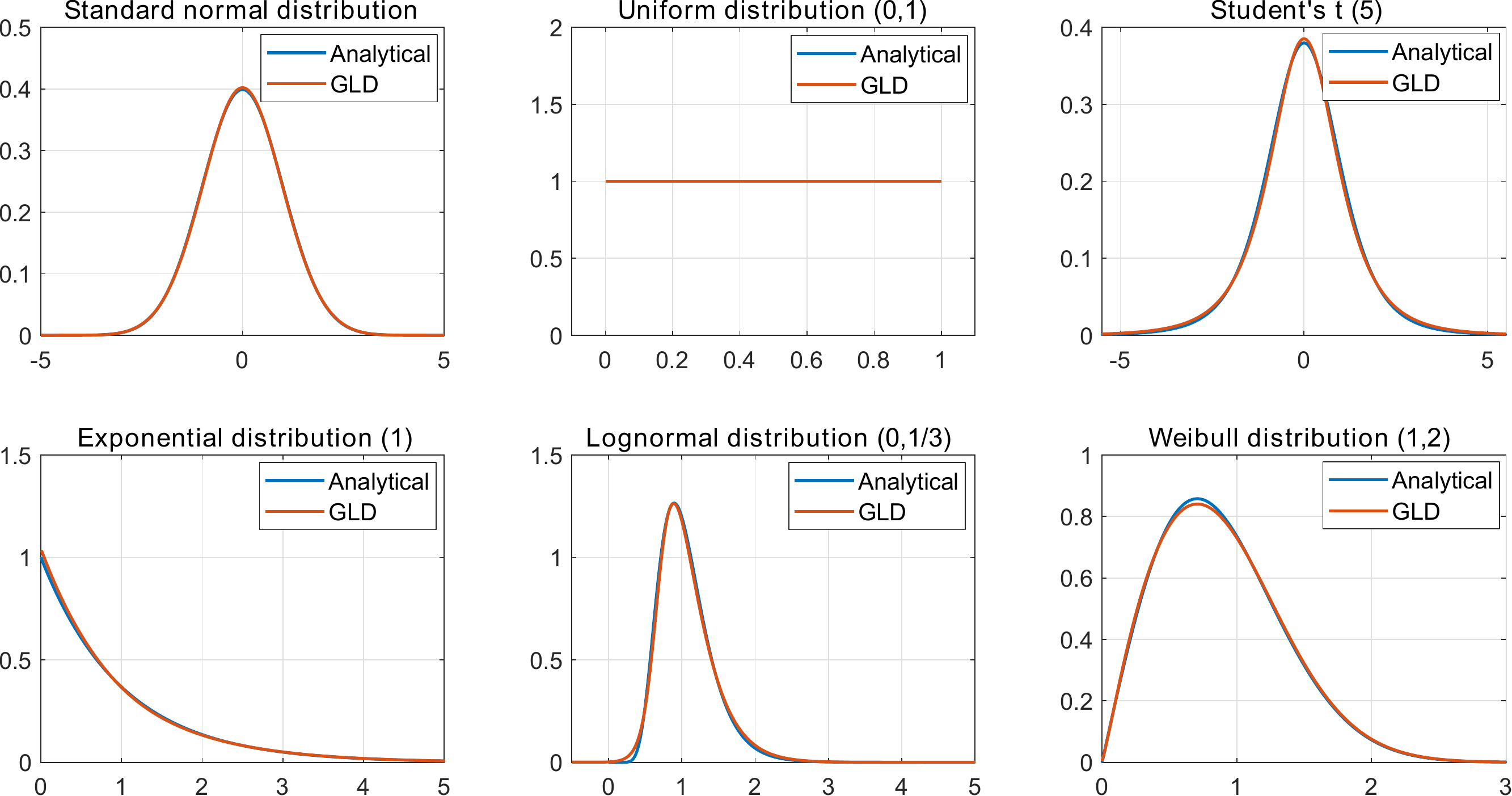}
	\caption{Visual comparison of GLD approximation of several common distributions.}
	\label{fig:GLD_approx}
\end{figure}

The generalized lambda distribution (GLD) is a highly flexible four-parameter probability distribution function designed to approximate most of the well-known parametric distributions \citep{Karian2000}. \Cref{fig:GLD_approx} illustrates how, with the proper choice of parameters, it can accurately approximate, normal, uniform, Student's $t$, exponential, lognormal, Weibull distributions, among others.
\par
Instead of providing a direct parametrization of the PDF, the GLD parametrizes the \emph{quantile function}, which is the inverse of the cumulative distribution function $Q = F^{-1}(u)$. Therefore, $Q$ is a non-decreasing function defined in $[0,1]$. In this paper, we consider the GLD of the Freimer-Kollia-Mudholkar-Lin (FKML) family \citep{Freimer1988}, which is defined as:
\begin{equation}\label{eq:FKML}
Q(u) = \lambda_1 + \frac{1}{\lambda_2}\left( \frac{u^{\lambda_3}-1}{\lambda_3} - \frac{(1-u)^{\lambda_4}-1}{\lambda_4}\right),
\end{equation}
where $\lambda_1$ is the location parameter, $\lambda_2$ is the scaling parameter, and $\lambda_3$ and $\lambda_4$ are shape parameters. To ensure valid quantile functions, it is only required that $\lambda_2$ be positive. 
\par
Parametrizing the quantile function is equivalent to modeling the inverse probability integral transform. More precisely, the random variable $Y$ with $Q$ as quantile function and the random variable $Q(U)$ with $U \sim \cu\left(0,1\right)$ follow the same distribution. Therefore, the PDF $f_{Y}(y)$ of a random variable $Y$ following a GLD can be calculated through a change of variables as follows:
\begin{equation}\label{eq:FKMLpdf}
f_Y(y) = \frac{f_U(u)}{Q^\prime(u)} =\frac{\lambda_2}{ u^{\lambda_3-1} + (1-u)^{\lambda_4-1}} \mathbbm{1}_{[0,1]}(u), \text{ with } u = Q^{-1}(y), \\
\end{equation}
where $\mathbbm{1}_{[0,1]}$ is the indicator function. A closed form expression of $Q^{-1}$, and therefore of $f_Y$, is in general not available.
\par
\begin{figure}[htbp!]
	\centering
	\includegraphics[width=0.95\linewidth]{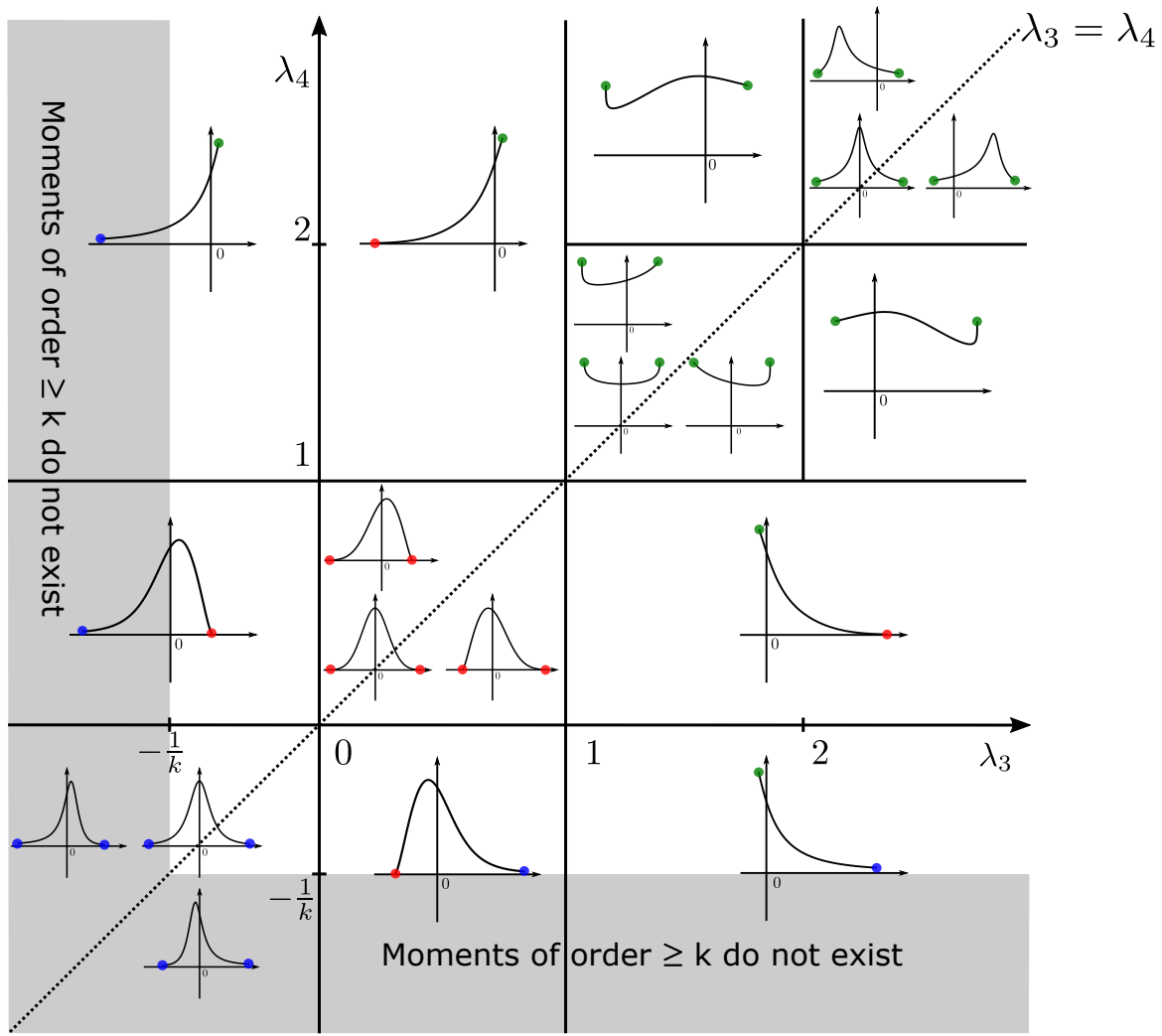}
	\caption{A graphical illustration of the shapes that can be represented by the FKML family of GLD as a function of $\lambda_3$ and $\lambda_4$. The values of $\lambda_1$ and $\lambda_2$ are set to 0 and 1, respectively. The dotted line is $\lambda_3 = \lambda_4$, which produces symmetric PDFs. The blue points indicate that the PDF has infinite support in the marked direction. In contrast, both the red and green points denote the boundary points of the PDF support. The PDF $f_Y(y) = 0$ on the red dots, whereas $f_Y(y) = 1$ on the green ones.}
	\label{fig:FKML_shape}
\end{figure}
\par
\Cref{fig:FKML_shape} illustrates some PDF shapes of the FKML generalized lambda distributions in the $(\lambda_3,\lambda_4)$ plane. It shows that distributions which belong to this family can cover a wide range of shapes that are determined by $\lambda_3$ and $\lambda_4$. For example, $\lambda_3 = \lambda_4$ produces symmetric PDFs, and $\lambda_3, \lambda_4<1$ yields bell-shaped distributions. Importantly, $\lambda_3$ and $\lambda_4$ control the support and the tail properties of the resulting PDF. The distribution has lower infinite support for $\lambda_3 \leq 0$ and upper infinite support for $\lambda_4 \leq 0$. In contrast, $\lambda_3>0$ implies that the PDF support is left-bounded and $\lambda_4>0$ corresponds to right-bounded distributions. More precisely, the support of the PDF, denoted by $\support{f_Y(y)} = [B_l,B_u]$, can be derived from \refEq{eq:FKML} as follows:
\begin{equation}
\begin{split}\label{eq:Bounds}
B_l\left(\lambda_1,\lambda_2,\lambda_3\right) &= \begin{cases}
-\infty, &\lambda_3 \leq 0 \\
\lambda_1 - \frac{1}{\lambda_2 \lambda_3}, &\lambda_3 > 0
\end{cases}, \\
B_u\left(\lambda_1,\lambda_2,\lambda_4\right) &= \begin{cases}
+\infty, &\lambda_4 \leq 0 \\
\lambda_1 + \frac{1}{\lambda_2 \lambda_4}, &\lambda_4 > 0
\end{cases}.
\end{split}
\end{equation}

\subsection{Estimation of \texorpdfstring{$\ve{\lambda}$}{TEXT}}
\label{sec:Pfit}

Many estimation methods have been proposed to fit a generalized lambda distribution to data \citep{Chalabi2010}. \cite{Karian2010,Corlu2016} compared different methods through exhaustive Monte Carlo simulations with various test cases. All of the estimators show comparable performance, and none of them is shown to always outperform the others. The performance depends on the shape of the true distribution, the sample size and the goodness-of-fit criterion used for comparison. In this paper, we choose to apply the method of moments that relies on matching the four moments: mean, variance, skewness, and kurtosis \citep{Lakhany2000} and the maximum likelihood estimation \citep{Su2007}.
\par
\subsubsection{Method of moments}
Following \refEq{eq:FKMLpdf}, the expectation of any function $g(Y)$ can be calculated as
\begin{equation}\label{eq:exp_func}
\Esp{g(Y)} = \Esp{g(Q(U))} = \int_{0}^{1} g(Q(u))du.
\end{equation}
Accordingly, the $k^{\text{th}}$ moment is given by
\begin{align*}
\Esp{Y^k} &= \int_{0}^{1}  \left(\lambda_1 + \frac{1}{\lambda_2}\left( \frac{u^{\lambda_3}-1}{\lambda_3} - \frac{(1-u)^{\lambda_4}-1}{\lambda_4}\right)\right)^k du \\
&=\int_{0}^{1} \left(\lambda_1 - \frac{1}{\lambda_2 \lambda_3} + \frac{1}{\lambda_2 \lambda_4} + \frac{1}{\lambda_2}\left( \frac{u^{\lambda_3}}{\lambda_3} - \frac{(1-u)^{\lambda_4}}{\lambda_4}\right)\right)^k du ,
\end{align*}
which is then simplified as
\begin{equation}\label{eq:GLD_mom}
\Esp{Y^k}=\int_{0}^{1} \left(c+\frac{1}{\lambda_2}s(u)\right)^k du, 
\end{equation}
where
\begin{align*}
c \eqdef \lambda_1 - \frac{1}{\lambda_2 \lambda_3} + \frac{1}{\lambda_2 \lambda_4}, \\
s(u) \eqdef \frac{u^{\lambda_3}}{\lambda_3} - \frac{(1-u)^{\lambda_4}}{\lambda_4}.
\end{align*}
To further elaborate \refEq{eq:GLD_mom}, we calculate
\begin{equation}\label{eq:GLD_MM_aux}
v_k = \int_{0}^{1} s(u)^k du = \sum_{j=0}^{k} \frac{(-1)^j}{\lambda_3^{k-j}\lambda_4^j}
\begin{pmatrix} k \\ j \end{pmatrix} \Betafun(\lambda_3(k-j)+1,\lambda_4j+1) ,
\end{equation}
where $\Betafun$ denotes the beta function. With the help of \refEq{eq:GLD_MM_aux}, \refEq{eq:GLD_mom} can be calculated through binomial expansions. As a result, the mean, variance, skewness, and kurtosis are given by (see details in \cite{Lakhany2000})
\begin{align}
\mu &= \Esp{Y} = \lambda_1 - \frac{1}{\lambda_2}\left(\frac{1}{\lambda_3+1} - \frac{1}{\lambda_4+1}\right), \label{eq:GLD_mean}\\
\sigma^2 &= \Esp{(Y-\mu)^2} = \frac{(v_2-v_1^2)}{\lambda_2^2}, \label{eq:GLD_var}\\
\delta &= \Esp{\left(\frac{(Y-\mu)}{\sigma}\right)^3} = \frac{v_3 - 3v_1v_2+2v_1^3}{(v_2-v_1^2)^{\frac{3}{2}}}, \label{eq:GLD_skew}\\
\kappa &=\Esp{\left(\frac{(Y-\mu)}{\sigma}\right)^4} = \frac{v_4-4v_1v_3+6v_1^2v_2-3v_1^4}{\left(v_2-v_1\right)^2}. \label{eq:GLD_kurt}
\end{align}
\par
The method of moments matches these four quantities to the associated empirical moments $\left(\hat{\mu},\hat{\sigma}^2,\hat{\delta},\hat{\kappa}\right)$ computed from the available sample set $\cy = \acc{y^{(1)}\ldots,y^{(N)}}$. Since $v_k$ is only a nonlinear function of $\lambda_3$ and $\lambda_4$, the skewness and kurtosis are also functions of only $\lambda_3$ and $\lambda_4$. Therefore, the fitting procedure first estimates $\lambda_3, \lambda_4$ solving \refEqs{eq:GLD_skew}{eq:GLD_kurt}, which can be replaced by an optimization problem shown in \refEq{eq:GLD_lam34}. The remaining parameters, namely $\lambda_1$ and $\lambda_2$, are then estimated directly from \refEqs{eq:GLD_mean}{eq:GLD_var}.
\begin{equation}\label{eq:GLD_lam34}
\left(\hat{\lambda}_3, \hat{\lambda}_4\right) = \arg\min_{\lambda_3,\lambda_4} (\delta(\lambda_3,\lambda_4) - \hat{\delta})^2 + (\kappa(\lambda_3,\lambda_4) - \hat{\kappa})^2.
\end{equation}
Note that for $\lambda_3 \leq -0.25$ or $\lambda_4 \leq -0.25$, the generalized lambda distribution has infinite fourth order moment as shown in \Cref{fig:FKML_shape} for $k=4$. Therefore, the method of moments only provides $\lambda_3>-0.25$ and $\lambda_4>-0.25$. 

\subsubsection{Maximum likelihood estimation}

Since the PDF of the generalized lambda distribution is not explicitly given, the negative log-likelihood function can only be evaluated numerically according to \refEq{eq:FKMLpdf}:
\begin{equation}\label{eq:logl}
l(\ve{\lambda}) = -\sum_{i=1}^{N} \log\left(\frac{\lambda_2}{u_i^{\lambda_3-1} + (1-u_i)^{\lambda_4-1}}\right) ,
\end{equation}
where
\begin{equation}\label{eq:nonl}
\text{where } u_i = Q^{-1}\left(y_{i}\right), \,\,\, y_{i} = Q(u_i) = \lambda_1 + \frac{1}{\lambda_2}\left(\frac{u_i^{\lambda_3}-1}{\lambda_3} - \frac{(1-u_i)^{\lambda_4}-1}{\lambda_4}\right). 
\end{equation}
\par
The maximum likelihood method estimates the distribution parameters by minimizing the negative log-likelihood defined in \refEq{eq:logl}:
\begin{equation}\label{eq:GLD_MLE}
\hat{\ve{\lambda}} = \arg\min_{\ve{\lambda}} l(\ve{\lambda}).
\end{equation}
For a sample set of size N, each likelihood function evaluation requires solving $N$ times the nonlinear equation \refEq{eq:nonl}. Consequently, the maximum likelihood estimation can be time-consuming for large data sets. To alleviate the computational burden, we propose the bisection method \citep{Burden2015} to efficiently solve \refEq{eq:nonl} using the property that $Q(u)$ is monotonic and defined in $[0,1]$.

\section{Polynomial chaos expansions}
\label{sec:PCE}

\subsection{Introduction}
A deterministic simulator is a function $\cm$ that maps a set of input parameters $\ve{x} =\left(x_1,x_2,\ldots,x_M\right)^T \in \cd_{\ve{X}} \subset\Rr^M$ to the output quantity of interest $y \in \Rr$. In the context of uncertainty quantification, the input parameters are modeled by a random vector $\ve{X}=\left(X_1,X_2,\ldots,X_M\right)^T$ described by its joint distribution $f_{\ve{X}}$ with support $\cd_{\ve{X}}$. Therefore, the uncertainty in the input variables propagates through the computational model to the output, which becomes a random variable denoted by $Y = \cm(\ve{X})$. 
\par
Under the assumption that $Y$ has finite variance, $Y$ belongs to the Hilbert space $\ch$ of square-integrable functions with respect to the following inner product:
\begin{equation}
\langle u,v\rangle_{\ch} = \Esp{u(\ve{X})v(\ve{X})} = \int_{\cd_{\ve{X}}} u(\ve{x})v(\ve{x})f_{\ve{X}}(\ve{x}) \D\ve{x}.
\end{equation}
If the joint distribution $f_{\ve{X}}$ satisfies certain conditions \citep{Ernst2012}, the set of multivariate polynomials is dense in $\ch$. Hence, $\ch$ is a separable Hilbert space admitting a polynomial basis $\acc{\psi_{\ve{\alpha}}(\cdot),\ve{\alpha} \in \Nn^M}$ which satisfies
\begin{equation}\label{eq:PCE_ortho}
\langle \psi_{\ve{\alpha}},\psi_{\ve{\beta}}\rangle_{\ch} = \delta_{\ve{\alpha}\ve{\beta}},
\end{equation}
with $\delta$ being the Kronecker symbol defined by $\delta_{\ve{\alpha}\ve{\beta}}=1$ if $\ve{\alpha}=\ve{\beta}$ and $\delta_{\ve{\alpha}\ve{\beta}}=0$ otherwise. Each component $\alpha_j$ of $\ve{\alpha}$ indicates the polynomial degree of $\psi_{\ve{\alpha}}$ in the variable $x_j$. As a results, $\cm$ can be represented by
\begin{equation}\label{eq:PCE}
\cm(\ve{X}) = \sum_{\ve{\alpha}\in \Nn^M}a_{\ve{\alpha}}\psi_{\ve{\alpha}}(\ve{X}),
\end{equation}
where $a_{\ve{\alpha}}$ is the coefficient associated to the basis function $\psi_{\ve{\alpha}}$. The construction of $\acc{\psi_{\ve{\alpha}}(\cdot),\ve{\alpha} \in \Nn^M}$ for arbitrary $f_{\ve{X}}(\ve{x})$ is in general difficult. In this study, we consider that $\ve{X}$ has mutually independent components. Thus, the joint distribution $f_{\ve{X}}$ is expressed as
\begin{equation}
f_{\ve{X}}(\ve{x}) = \prod_{j=1}^{M}f_{X_j}(x_j) .
\end{equation}
In this case, each basis function $\psi_{\ve{\alpha}}$ that fulfils \refEq{eq:PCE_ortho} can be obtained as the tensor product of univariate polynomials: 
\begin{equation}\label{eq:PCE_basis}
\psi_{\ve{\alpha}}(\ve{x}) = \prod_{j=1}^{M} \phi^{(j)}_{\alpha_j}(x_j),
\end{equation}
where $\acc{\phi^{(j)}_{\alpha_j},\alpha_j \in \Nn}$ are orthogonal polynomials with respect to the marginal distribution of $f_{X_j}$, \ie
\begin{equation}\label{eq:unPCE_ortho}
\Esp{\phi^{(j)}_{\alpha_j}(X_j)\cdot\phi^{(j)}_{\beta_j}(X_j)}=\delta_{\alpha_j\beta_j}.
\end{equation}
As a result, the problem of constructing $\psi_{\ve{\alpha}}$ is reduced to finding univariate orthogonal polynomials. Polynomials that are orthogonal with respect to some classical distributions, \eg normal, uniform, exponential, are listed in \cite{Xiu2002}. For arbitrary marginal distributions, orthogonal polynomials can be calculated numerically through the \emph{Stieltjes procedure} \citep{Gautschi2004}.

\subsection{Sparse PCE}
The spectral expansion in \refEq{eq:PCE} is an infinite series. In practice, truncation schemes must be adopted, which leads to approximating the computational model by a finite series defined by a finite multi-index subset $\caa \subset \Nn^M$. 
\begin{equation}
\cm(\ve{x}) \approx \cm^{PC}(\ve{x}) = \sum_{\ve{\alpha}\in \caa} a_{\ve{\alpha}}\psi_{\ve{\alpha}}(\ve{x})
\end{equation}
Once the set of candidates is selected, regression-based algorithms such as ordinary least squares \citep{Berveiller2006} can be applied to the data $\left(\ve{\cx},\cy\right) = \acc{\left(\ve{x}^{(i)},y^{(i)}\right), i = 1,\ldots,N}$ to build the surrogate model. Here $\ve{\cx}$ denotes the experimental design of the input variables, and $\cy$ are the associated model outputs. One common method to select $\caa$ is the full basis of degree $p$, which contains all the PC basis functions the degree of which is lower than a given value $p$. However, it is well known that the classical ``full'' PC approximation suffers from the curse of dimensionality \citep{SudretJCP2011}, due to the quick increase of the basis size with increasing input dimension or polynomial degree. To overcome this problem, sparse polynomial chaos expansions have been proposed, which select only the most important basis functions among a candidate set \citep{BlatmanPEM2010,SudretJCP2011}, before ordinary least squares are used to compute the coefficients. In the present work, we use the hybrid-LAR algorithm \citep{UQdoc_13_104} implemented in the open source software UQLab \citep{MarelliUQLab2014} for building sparse PCE. The selection procedure of the algorithm is based on \emph{least angle regression} (LAR) \citep{Efron2004}.  
\par
In the sequel, we will combine PCE with the local inference of generalized lambda distributions on each point of the experimental design with replications.

\section{Infer-and-Fit algorithm and joint modeling}
\label{sec:JM}

\subsection{Introduction}
We assume that the response PDF of the stochastic simulator for a given input realization $\ve{x}$ follows a generalized lambda distribution, with distribution parameters $\ve{\lambda} = \left(\lambda_1,\lambda_2,\lambda_3,\lambda_4\right)^T$ that are functions of $\ve{x}$:
\begin{equation} \label{eq:condGLD}
Y (\ve{x}) \sim \GLD\left(\lambda_1(\ve{x}),\lambda_2(\ve{x}),\lambda_3(\ve{x}),\lambda_4(\ve{x})\right).
\end{equation}
\par
Under appropriate assumptions discussed in \Cref{sec:PCE}, each component of $\ve{\lambda}(\ve{x})$ admits a PC representation. For the FKML family, $\lambda_2(\ve{x})$ is required to be positive (see \Cref{sec:GLD}), and thus the associated PC approximation is built on the natural logarithm $\log\left(\lambda_2(\ve{x})\right)$. In a nutshell, $\ve{\lambda}(\ve{x})$ are decomposed as
\begin{align}
\lambda_s\left(\ve{x}\right) &\approx \lambda^{\PC}_s\left(\ve{x};\ve{a}\right) = \sum_{\ve{\alpha}\in \caa_s} a_{s,\ve{\alpha}}\psi_{\ve{\alpha}}(\ve{x}), \quad s = 1,3,4 \label{eq:lamPCE}\\
\lambda_2\left(\ve{x}\right) &\approx \lambda^{\PC}_2\left(\ve{x};\ve{a}\right) = \exp \left(\sum_{\ve{\alpha}\in \caa_2} a_{2,\ve{\alpha}}\psi_{\ve{\alpha}}(\ve{x}) \right), \label{eq:lamPCE_lam2}
\end{align}
\noindent
where $\ve{\lambda}^{\PC}\left(\ve{x};\ve{a}\right)$ are the PC approximations of the unknown functions $\ve{\lambda}(\ve{x})$. The truncation sets $\acc{\caa_s,s=1,2,3,4}$ are to be defined, and the coefficients $a_{s,\ve{\alpha}}$ are the model parameters to be estimated from the samples. For the purpose of clarification, we explicitly express the model parameters $\ve{a}$ in the surrogate model $\ve{\lambda}^{\PC}\left(\ve{x};\ve{a}\right)$ so as to emphasize that $\ve{a}$ are unknown and need to be estimated from the data.

\subsection{Infer-and-Fit algorithm}
To account for the intrinsic randomness, the stochastic simulator is repeatedly run $R$ times for each point $\ve{x}^{(i)}$ of the experimental design $\ve{\cx}$, and the associated output is denoted by $\cy^{(i)} = \acc{y^{(i,1)},y^{(i,2)},\ldots,y^{(i,R)}}$, where the upper index $(i,r)$ refers to the output of the $r^{th}$ replication for the $i^{th}$ set of input parameters in the experimental design. 
\par
Following \cite{Moutoussamy2015,Browne2016}, one straightforward way to build a surrogate model is the Infer-and-Fit algorithm presented in \Cref{alg:twostep}.
\par
\begin{figure}[htb]
	\centering
	\begin{minipage}{.89\linewidth}
		\begin{algorithm}[H]
			\caption{Infer-and-Fit algorithm}
			\label{alg:twostep}
			\begin{algorithmic}[1]
				\FOR{$i \gets 1,N$ }
				\STATE $\hat{\ve{\lambda}}^{(i)} \gets \hat{\ve{\lambda}}\left(\cy^{(i)}\right)$
				\ENDFOR
				\STATE $\hat{\ve{\Lambda}} \gets \left(\hat{\ve{\lambda}}^{(1)},\hat{\ve{\lambda}}^{(2)} ,\ldots,\hat{\ve{\lambda}}^{(N)}\right)^T$
				\STATE{$\ve{\lambda}^{\PC}(\ve{x};\tilde{\ve{a}}) \gets \LAR\left(\ve{\cx},\hat{\ve{\Lambda}}\right)$}
			\end{algorithmic}
		\end{algorithm}
	\end{minipage}
\end{figure}
\par
Function $\hat{\ve{\lambda}}(\cdot)$ in the second line of \Cref{alg:twostep} denotes an estimator of the distribution parameters based on the replications (see \Cref{sec:Pfit}), and $\LAR$ in the last line is the algorithm \citep{SudretJCP2011} used to build sparse PCE for $\ve{\lambda}(\ve{x})$. 
\par
\Cref{alg:twostep} consists of two main steps. The first step is used to capture the intrinsic stochasticity through replications. More precisely, this inference step aims at providing an estimate $\hat{\ve{\lambda}}^{(i)}$ of the distribution parameters $\ve{\lambda}\left(\ve{x}^{(i)}\right)$ for each point of the experimental design $\ve{\cx}$. The second step independently builds four surrogate models for the distribution parameters, based on the estimated parameters at discrete points of the experimental design. For the local inference in the first step, we test both the method of moments and the maximum likelihood estimation (a detailed comparison is presented in \Cref{sec:examples}). Besides, in the second step, we choose to use the hybrid-LAR for sparse PCE constructions, but \Cref{alg:twostep} is not bounded to this choice: any other regression methods such as ordinary least squares \citep{Berveiller2006}, orthogonal matching pursuit \citep{Tropp2007}, etc. can be used equivalently.
\par
In practice, the estimator $\hat{\ve{\lambda}}^{(i)}$ is calculated from replications of finite size due to finite computational budget, and it is subject to noise. Consequently, the choice of the regression setting for building sparse PCE is advantageous because of its robustness to noise \citep{Torre2019}. However, the generalized lambda distribution is rather flexible, so that a few samples cannot guarantee an accurate estimation \citep{Corlu2016}, and none of the existing methods have been proved to produce unbiased estimators. If a consistent bias is present in the estimation, the use of regression algorithms cannot filter it out. Moreover, the four parameters of the GLD, considered as functions of the input variables, are approximated by four PCE built independently. As a result, the Infer-and-Fit algorithm qualitatively requires many replications $R$ to achieve a good estimate (quantitative results are shown in \Cref{sec:examples}). 
\par
The two separate steps of \Cref{alg:twostep} may be seen as two successive, independent optimization problems. First, the four parameters of the GLD are optimally fitted for each point $\ve{x}^{(i)} \in \ve{\cx}$, leading to $\hat{\ve{\Lambda}}$. Second, coefficients of the PCE of each parameter $\lambda_s(\ve{x})$ are optimized based on the data collected in $\hat{\ve{\Lambda}}$, so as to minimize a mean squared error. Intuitively, it appears that these two successive optimizations are suboptimal. We propose to complement the Infer-and-Fit algorithm with a subsequent joint optimization.

\subsection{Joint PCE-GLD fitting}
To reduce the computational cost associated with the need for a large number of replications, we propose a similar approach as that of generalized linear models \citep{McCullagh1989}. In this joint modeling method, PC coefficients $\ve{a}$ of the four $\lambda_s$'s are calibrated from the original data $(\ve{\cx},\ve{\cy})$ through a maximum likelihood estimation, instead of being calibrated from the local estimates $\hat{\ve{\Lambda}}= \acc{\hat{\ve{\Lambda}}^{(1)},\ldots,\hat{\ve{\Lambda}}^{(N)}}$, as shown in \Cref{alg:twostep}.
\par
To form such an estimator, a deeper insight into the nature of stochastic simulators is necessary. Running once the stochastic simulator for $\ve{x}$, the output value is a realization of the random variable $Y(\ve{x})$, which can also be written as $Y \mid \ve{X} = \ve{x}$. As a result, the stochastic simulator can be regarded as a \emph{conditional sampler} with the response PDF $f_{Y\mid\ve{X}}\left(y\mid\ve{x}\right)$. Therefore, we can write the joint distribution of $(\ve{X},Y)$ as $f_{\ve{X},Y}(\ve{x},y) = f_{Y\mid\ve{X}}\left(y\mid\ve{x}\right)f(\ve{x})$. The GLD surrogate provides an approximation $f_{Y\mid\ve{X}}\left(y\bigr\rvert \ve{\lambda}^{\PC}(\ve{x};\ve{a}) \right)$ to the conditional PDF. Therefore, the joint PDF of the GLD model is $f_{\ve{X},Y}(\ve{x},y;\ve{a}) = f_{\ve{X}}(\ve{x})f_{Y\mid\ve{X}}\left(y\bigr\rvert\ve{\lambda}^{\PC}(\ve{x};\ve{a})\right)$.
\par
Minimizing the Kullback-Leibler divergence between $f_{\ve{X},Y}(\ve{x},y)$ and $f_{\ve{X},Y}(\ve{x},y;\ve{a})$ gives an appropriate approximation of the GLD surrogate to the underlying true model:
\begin{equation}\label{eq:defa0}
\ve{a}_0 = \arg\min_{\ve{a}} D_{KL}\left(f_{\ve{X},Y}\left(\ve{x},y\right) \| f_{\ve{X},Y}\left(\ve{x},y ; \ve{a}\right)\right),
\end{equation}
where:
\begin{equation}
\begin{split}\label{eq:defKL}
D_{KL}\left(f_{\ve{X},Y}\left(\ve{x},y\right) \| f_{\ve{X},Y}\left(\ve{x},y ; \ve{a}\right)\right) &= \int f_{\ve{X},Y}\left(\ve{x},y \right) \log\left(\frac{f_{\ve{X},Y}\left(\ve{x},y\right)}{f_{\ve{X},Y}\left(\ve{x},y ; \ve{a}\right)}\right) \D\ve{x} \D y \\
&= -\int f_{\ve{X},Y}\left(\ve{x},y\right) \log\left(f_{\ve{X},Y}\left(\ve{x},y ; \ve{a}\right)\right) \D\ve{x} \D y +{\rm const.}\\
&= -\int f_{\ve{X},Y}\left(\ve{x},y\right) \log\left(f_{Y \mid \ve{X}}\left(y \mid \ve{x}; \ve{a}\right) f_{\ve{X}}(\ve{x})\right) \D\ve{x} \D y  +{\rm const.}
\end{split}
\end{equation}
\par
Since $f_{\ve{X}}$ does not contain the model parameters $\ve{a}$, \refEq{eq:defa0} can be further simplified as
\begin{align}
\ve{a}_0= \arg\min_{\ve{a}} -\int f_{\ve{X},Y}\left(\ve{x},y\right) \log\left(f_{Y \mid \ve{X}}\left(y \mid \ve{\lambda}(\ve{x};\ve{a})\right)\right) \D\ve{x} \D y. \label{eq:defl}
\end{align}
Thus the model parameters $\ve{a}_0$ can be obtained by minimizing the following function:
\begin{equation}
l(\ve{a}) \eqdef -\Espe{\ve{X},Y}{\log\left(f_{Y \mid \ve{X}}\left(Y \Bigr\rvert \ve{\lambda}^{\PC}(\ve{X};\ve{a})\right)\right)}.\label{eq:KL}
\end{equation}
Note that if the true underlying model can be expressed as the form $f_{Y\mid\ve{X}}\left(y\bigr\rvert \ve{\lambda}^{\PC}\left(\ve{x};\ve{a}_{\rm true}\right) \right)$, $\ve{a}_0$ from \refEq{eq:defl} guarantees that $f_{Y\mid\ve{X}}\left(y\bigr\rvert \ve{\lambda}^{\PC}(\ve{x};\ve{a}_0) \right)$ is the same as that of the true model $\forall \ve{x} \in \cd_{\ve{x}}$. 
\par
To estimate $\ve{a}_0$, the expectation in \refEq{eq:KL} is replaced by some estimator. In most cases, a sample based empirical average $\frac{1}{N}\sum_{i=1}^{N} \log\left(f_{Y\mid\ve{X}}\left(y^{(i)} \Bigr\rvert \ve{\lambda}^{\PC}\left(\ve{x}^{(i)}; \ve{a}\right)\right)\right)$ is used, where $\acc{\left(\ve{x}^{(i)},y^{(i)}\right)}^{N}_{i = 1}$ are drawn independently from the joint distribution $f_{\ve{X},Y}\left(\ve{x},y\right)$. For a given $r$, $\acc{\left(\ve{x}^{(i)},y^{(i,r)}\right)}^{N}_{i = 1}$ are independent samples, and thus it is natural to consider the estimator
\begin{equation}
\hat{l}^{(r)}(\ve{a}) = \frac{1}{N}\sum_{i=1}^{N} -\log\left(f_{Y\mid\ve{X}}\left(y^{(i,r)} \Bigr\rvert \ve{\lambda}^{\PC}\left(\ve{x}^{(i)}; \ve{a}\right)\right)\right)
\end{equation}
to replace the expectation in \refEq{eq:KL}. Note that $\acc{\hat{l}^{(r)}(\ve{a})}^{R}_{r = 1}$ are unbiased estimators of $l(\ve{a})$. Therefore, the following estimator $\hat{l}(\ve{a})$ is also unbiased:
\begin{equation}\label{eq:estim}
\hat{l}(\ve{a})=\frac{1}{R}\sum_{r=1}^{R} \hat{l}^{(r)}(\ve{a}).
\end{equation}
For a given $\ve{a}$, $\acc{\hat{l}^{(r)}(\ve{a})}^{R}_{r = 1}$ have the same variance $\sigma^2(\ve{a})$, because they are the same estimator applied to different samples $\acc{\left(\ve{x}^{(1)},y^{(1,r)}\right),\ldots,\left(\ve{x}^{(N)},y^{(N,r)}\right)}$, generated by the same scheme and indexed by $r$. Nevertheless, these estimators of $l(\ve{a})$ are not mutually independent due to the presence of replications. Hence, the variance of $\hat{l}(\ve{a})$ is calculated as follows:
\begin{equation}
\begin{split}
\Var{\hat{l}(\ve{a})} &= \Var{\frac{1}{R}\sum_{r=1}^{R} \hat{l}^{(r)}(\ve{a})} \\
& = \frac{1}{R^2} \left(\sum_{r=1}^{R} \Var{\hat{l}^{(r)}(\ve{a})} + \sum_{r_1=1}^{R}\sum_{r_2 \neq r_1} \Cov{\hat{l}^{(r_1)}(\ve{a}),\hat{l}^{(r_2)}(\ve{a})} \right).
\end{split}
\end{equation}
Using the Cauchy-Schwartz inequality, we have
\begin{equation}
\begin{split}
\Var{\hat{l}(\ve{a})}& \leq \frac{1}{R^2} \left(\sum_{r=1}^{R} \Var{\hat{l}^{(r)}(\ve{a})} + \sum_{r_1=1}^{R}\sum_{r_2 \neq r_1} \sqrt{\Var{\hat{l}^{(r_1)}(\ve{a})}} \sqrt{\Var{\hat{l}^{(r_2)}(\ve{a})}} \right) \\
& = \frac{1}{R^2}\left(R \cdot \sigma^2(\ve{a}) + R(R-1)\cdot \sigma^2(\ve{a})\right) = \sigma^2(\ve{a}).
\end{split}
\end{equation}
The inequality becomes an equality if and only if $\hat{l}^{(r_1)}(\ve{a})$ is an affine function of $\hat{l}^{(r_2)}(\ve{a})$. In the context of stochastic simulators, $\hat{l}^{(r_1)}(\ve{a})$ is not a deterministic function of $\hat{l}^{(r_2)}(\ve{a})$. Therefore, for a given $\ve{a}$, $\hat{l}(\ve{a})$ has less variance than any estimator of the set $\acc{\hat{l}^{(r)}(\ve{a})}^{R}_{r = 1}$, so it is a better estimator in terms of variance. Replacing the expectation in \refEq{eq:KL} by $\hat{l}(\ve{a})$, we end up with a new estimator: 
\begin{equation}\label{eq:optim}
\hat{\ve{a}} = \arg\min_{\ve{a}}\hat{l}\left(\ve{a}\right),
\end{equation}
where
\begin{equation}\label{eq:nloglh}
\hat{l}\left(\ve{a}\right) \eqdef  \sum_{i}^{N}\sum_{r}^{R}-\log\left( 
f_{Y \mid X}\left(y^{(i,r)} \Bigr\rvert \ve{\lambda}^{\PC}\left(\ve{x}^{(i)};\ve{a}\right)\right)\right). 
\end{equation}
This estimator by itself does not produce sparsity in the PC representations, meaning that the basis functions for $\ve{\lambda}^{\PC}(\ve{x};\ve{a})$ should be predefined before optimizing \refEq{eq:optim}. To this end, we first exploit \Cref{alg:twostep} to identify a sparse truncation scheme $\acc{\caa_s, s = 1,\ldots,4}$ for each component of $\ve{\lambda}$ in terms of the input vector $\ve{x}$. Then, we keep this representation and optimize the associated coefficients over the $R \times N$ data points globally (by joint likelihood maximization \refEq{eq:optim}), instead of separately. Therefore, this new procedure, which is summarized in \Cref{alg:joint}, can be considered as a \emph{refinement} of \Cref{alg:twostep}, which is expected to improve the surrogate quality with respect to the number of available replications.
\par
\begin{figure}[htb]
	\centering
	\begin{minipage}{.89\linewidth}
		\begin{algorithm}[H]
			\caption{Joint PCE-GLD fitting}
			\label{alg:joint}
			\begin{algorithmic}[1]
				\STATE{Apply \Cref{alg:twostep} to get the sparse PCE truncation schemes $\acc{\caa_s, s = 1,\ldots,4}$, and the associated coefficients $\tilde{\ve{a}}$}
				\STATE{ $ \hat{\ve{a}} \leftarrow \arg\min_{\ve{a}}\hat{l}\left(\ve{a}\right)$,
					where $\hat{l}\left(\ve{a}\right)$ is defined in \refEq{eq:nloglh} and
					\begin{align}
					\lambda_s^{\PC}\left(\ve{x};\ve{a}\right) &= \sum_{\ve{\alpha}\in \caa_s} a_{s,\ve{\alpha}}\psi_{\ve{\alpha}}(\ve{x}) \,\,\, s = 1,3,4 \label{eq:lamPCE_2}\\
					\lambda_2^{\PC}\left(\ve{x};\ve{a}\right) &= \exp \left(\sum_{\ve{\alpha}\in \caa_2} a_{2,\ve{\alpha}}\psi_{\ve{\alpha}}(\ve{x}) \right) \label{eq:lamPCE_lam2_2}
					\end{align}} 
			\end{algorithmic}
		\end{algorithm}
	\end{minipage}
\end{figure}
\par
\begin{figure}[!htbp]
	\centering
	\includegraphics[width=0.65\linewidth]{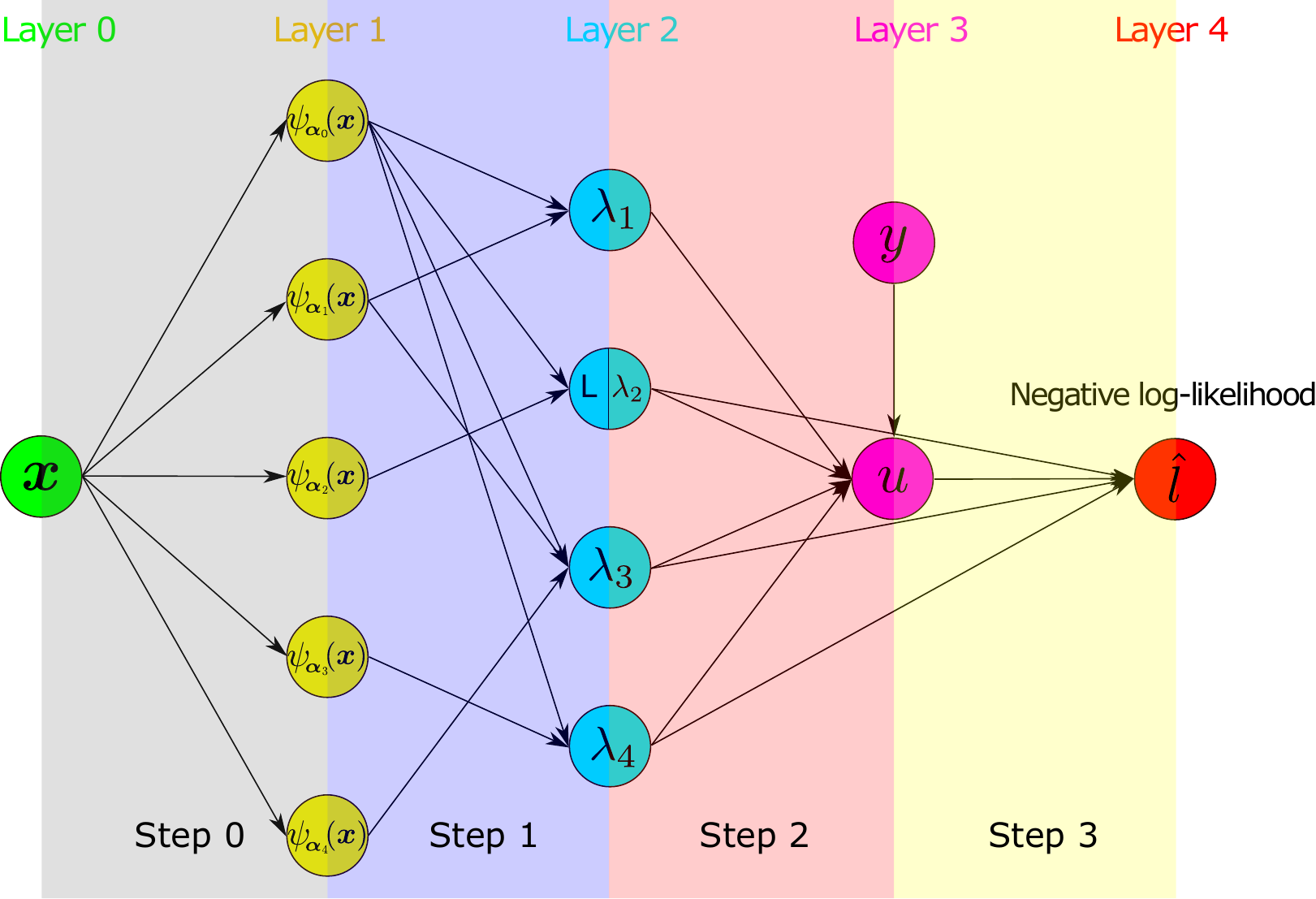}
	\caption{Flow chart of the negative log-likelihood calculation}
	\label{fig:network}
\end{figure}
\par
In the second step of \Cref{alg:joint}, the log-likelihood $\hat{l}(\ve{a})$ needs to be evaluated with given PC coefficients $\ve{a}$ for each data point $\left(\ve{x}^{(i)},y^{(i,r)}\right)$. The computation details are illustrated in \Cref{fig:network} and described here. The preliminary step (referred to as Step 0 in \Cref{fig:network}) evaluates the basis functions $\acc{\psi_{\ve{\alpha}},\ve{\alpha}\in \caa_s}$ at all $\ve{x}^{(i)} \in \cx$. Step 1 calculates the distribution parameters $\ve{\lambda}^{(i)} = \ve{\lambda}^{\PC}\left(\ve{x}^{(i)};\ve{a}\right)$ according to \refEqs{eq:lamPCE_2}{eq:lamPCE_lam2_2}, in which the model parameters $\ve{a}$ are used. The two layers involved in this step (Layer 1 and Layer 2 in \Cref{fig:network}) are not fully connected because the sparse basis sets are independently selected for each components of $\ve{\lambda}^{\PC}\left(\ve{x};\ve{a}\right)$ in \Cref{alg:twostep}. Step 2 solves the nonlinear equation $u_{i,r} = Q^{-1}\left(y^{(i,r)}\right)$, where the current values of $\ve{\lambda}^{(i)}$'s are used, see \refEq{eq:FKML}. The nonlinear equation is explicitly written as
\begin{equation}\label{eq:nl}
y^{(i,r)} = \lambda^{(i)}_1 + \frac{1}{\lambda^{(i)}_2}\left( \frac{u_{i,r}^{\lambda^{(i)}_3}-1}{\lambda^{(i)}_3} - \frac{(1-u_{i,r})^{\lambda^{(i)}_4}-1}{\lambda^{(i)}_4}\right).
\end{equation}
Eventually, Step 3 computes the negative log-likelihood using \refEq{eq:FKMLpdf}. More precisely, we have 
\begin{equation}
-\log\left(f_{Y \mid \ve{X}}\left(y^{(i,r)} \Bigr\rvert \ve{\lambda}^{(i)}\right)\right) = \log\left(\frac{u_{i,r}^{\lambda_3^{(i)}-1} + (1-u_{i,r})^{\lambda_4^{(i)}-1}}{\lambda_2^{(i)}}\right)\label{eq:lh}.
\end{equation}
\par
The optimization problem in the second step of \Cref{alg:joint} is not only highly nonlinear but also subject to complex constraints. As discussed in \Cref{sec:GLD}, the FKML family can produce PDFs with bounded support (see \refEq{eq:Bounds}), which implies that the negative log-likelihood function will take value $+\infty$ if the data are outside the support. To avoid numerical issues, constraints need to be introduced, and the complete optimization problem becomes
\begin{align}
\hat{\ve{a}} &= \arg\min_{\ve{a}} \hat{l}(\ve{a}) \label{eq:optimization}\\
\text{such that} \,\,\, \forall i \,\,\, 
&\begin{cases}
B_l\left(\lambda_1^{\PC}\left(\ve{x}^{(i)};\ve{a}\right),\lambda^{\PC}_2\left(\ve{x}^{(i)};\ve{a}\right),\lambda^{\PC}_3\left(\ve{x}^{(i)};\ve{a}\right)\right) \leq \min_r y^{(i,r)} \\
B_u\left(\lambda^{\PC}_1\left(\ve{x}^{(i)};\ve{a}\right),\lambda^{\PC}_2\left(\ve{x}^{(i)};\ve{a}\right),\lambda^{\PC}_4\left(\ve{x}^{(i)};\ve{a}\right)\right) \geq \max_r y^{(i,r)}
\end{cases}, \label{eq:constraints}
\end{align}
where $B_l$ and $B_u$ are computed from \refEq{eq:Bounds}.
\par
In general, the fact that the negative log-likelihood function can reach $+\infty$ is not a problem because \refEq{eq:optimization} is a minimization problem. Therefore, we can always treat the optimization problem as unconstrained. However, numerical issues can occur when applying unconstrained derivative-based algorithms. For this reason, we choose to use the derivative-based algorithm \emph{trust region without constraints} \citep{Steihaug1983} in the first place. If it does not converge, which implies that some constraints are activated, the constrained (1+1)-CMA-ES algorithm \citep{Arnold2012} available in UQLab \citep{UQdoc_13_201} is used instead.
\par
For derivative-based algorithms, the choice of a relevant starting point is important to ensure convergence. In the proposed method, we use the coefficients resulting from the Infer-and-Fit algorithm as the starting point, namely $\tilde{\ve{a}}$. However, $\tilde{\ve{a}}$ is generally not guaranteed to be feasible. If it violates the inequality conditions in \refEq{eq:constraints}, additional operations are necessary to have a feasible starting point, see details in \Cref{sec:feasible}.
\par
When applying derivative-based optimizers to solve \refEq{eq:optimization}, using finite difference to calculate gradients would be time-consuming and inaccurate. This is because the likelihood (\refEq{eq:nloglh}) is expensive to evaluate, especially considering that $NR$ nonlinear equations (\refEq{eq:nl}) that need to be solved. To alleviate the computational burden, we derived analytical expressions of the derivatives through implicit differentiations of \refEqs{eq:FKML}{eq:FKMLpdf} and the chain rule (see \Cref{sec:Deriv} for details). Besides, the Hessian matrix (second order derivatives of $\hat{l}$ with respect to $\ve{a}$) has also been derived. As a result, each iteration of the trust region algorithm only evaluates once the likelihood function $\hat{l}(\ve{a})$.

\section{Analytical examples}
\label{sec:examples}

In this section, we investigate the performance of the Infer-and-Fit and of the joint modeling algorithm using two analytical examples. The examples are built such that the PDF of $Y(\ve{x})$ is known but does not follow the generalized lambda distribution, so as to test the flexibility of the proposed approaches. As an inference tool for the first algorithm, we apply both the method of moments and the maximum likelihood estimation to get the values $\hat{\ve{\lambda}}^{(i)}$ for each $\ve{x}^{(i)} \in \ve{\cx}$. The associated surrogate models built from the Infer-and-Fit algorithm are respectively denoted by \GMM\ and \GMLE. Similarly, the joint PCE-GLD algorithm provides another two models denoted by \GJMM\ and \GJMLE. Note that when building these two joint models following \Cref{alg:joint}, both of them rely on solving the optimization problem in \refEq{eq:optimization}. However, results are not identical because the sparse truncation sets $\acc{\caa_s, s=1,\ldots,4}$ for $\ve{\lambda}^{\PC}(\ve{x};\ve{a})$ as well as the starting points for the optimization generally differ. 
\par
The error measure between the emulated PDF and the true one is computed using the Hellinger distance, which is then averaged over all possible $\ve{x}$. More precisely, we define
\begin{equation}\label{eq:Rlevel1} 
\epsilon = \Espe{\ve{X}}{d_{\text{HD}}\left(f_{Y\mid\ve{X}}(y\mid\ve{X}),f_{Y\mid\ve{X}}\left(y\mid\ve{\lambda}^{\PC}\left(\ve{X};\hat{\ve{a}}\right)\right)\right)}. 
\end{equation}
It is reminded that the Hellinger distance between two continuous PDFs $p$ and $q$ reads
\begin{equation}
\begin{split}
d_{\text{HD}}\left(p(y),q(y)\right) &= \frac{1}{\sqrt{2}} \norm{\sqrt{p(y)} - \sqrt{q(y)}}_2 \\
&= \sqrt{\frac{1}{2}\int_{-\infty}^{+\infty}\left(\sqrt{p(y)} - \sqrt{q(y)}\right)^2\D y} = \sqrt{1-\int_{-\infty}^{+\infty}\sqrt{p(y)q(y)}\D y}.\label{eq:HD}
\end{split}
\end{equation}
\par
Another natural choice for measuring the distance between two PDFs would have been the KL divergence. However, $D_{KL}(p\|q)$ tends to $+\infty$ if $\suppt(p) \setminus \suppt(q)$ has non zero probability with respect to $p$, which is not suitable for the comparison.
\par
In practice, the integral in \refEq{eq:HD} is computed using numerical integration. In this paper, we restrict the integral interval from $(-\infty,+\infty)$ to $[Q^{0.001}_p,Q^{0.999}_p] \cup [Q^{0.001}_q,Q^{0.999}_q]$, where $Q^{0.001}_p$ and $Q^{0.999}_p$ denote the $0.1\%$ and $99.9\%$ quantiles of a random variable having $p$ as PDF (similar notations are used for $q$). Note that this is feasible here because the two densities in \refEq{eq:Rlevel1} we want to compare have analytical expressions for the specific examples handled.
\par
To calculate the expectation in \refEq{eq:Rlevel1}, quasi Monte Carlo simulation is used with $N_{\rm test} = 1,000$ samples generated by the Sobol' sequence \citep{Sobol1967} in the input space. The Sobol' sequence sampler is also used to draw the experimental design (ED). To study the performance of the proposed methods, data are generated for various combinations of the experimental design size $N$ and the amount of replications $R$ per ED point. Each scenario is run 100 times with independent experimental designs to account for statistical uncertainty. Error estimates for each scenario $(N,R)$ are thus represented by box plots.

\subsection{Example 1: a one-dimensional simulator}

The first example is defined as follows:
\begin{equation}\label{ex:logn}
Y(X,\omega) = \sin\left(\frac{2\pi}{3}X + \frac{\pi}{6}\right) \cdot \left(Z_1(\omega) \cdot Z_2(\omega) \right)^{\cos(X)},
\end{equation}
where $X \sim \cu(0,1)$ is the input parameter, and $Z_1(\omega) \sim \mathcal{LN}(0,0.25)$ and $Z_2(\omega) \sim \mathcal{LN}(0,0.5)$ are latent variables following lognormal distributions. Under this definition, $Y(x,\omega)$ follows a lognormal distribution $\cl\cn(\ell(x),\zeta(x))$ with $\ell(x) = \log\left(\sin\left(\frac{2\pi}{3}x + \frac{\pi}{6}\right)\right)$ and $\zeta(x)=\sqrt{\frac{3}{8}}\cos(x)$, $x \in [0,1]$. As mentioned in \Cref{sec:GLD}, the lognormal distribution, which is widely used in engineering, can be approximated by the generalized lambda distribution. The nonlinearity in its parameters leads to nonlinear functions of $\ve{\lambda}(\ve{x})$ in the GLD approximation, and thus the PC representations $\ve{\lambda}^{\PC}(\ve{x})$ are also nonlinear.
\par
\begin{figure}[!htbp]
	\centering
	\subfigure[Generated data]{\includegraphics[height=.33\linewidth, keepaspectratio]{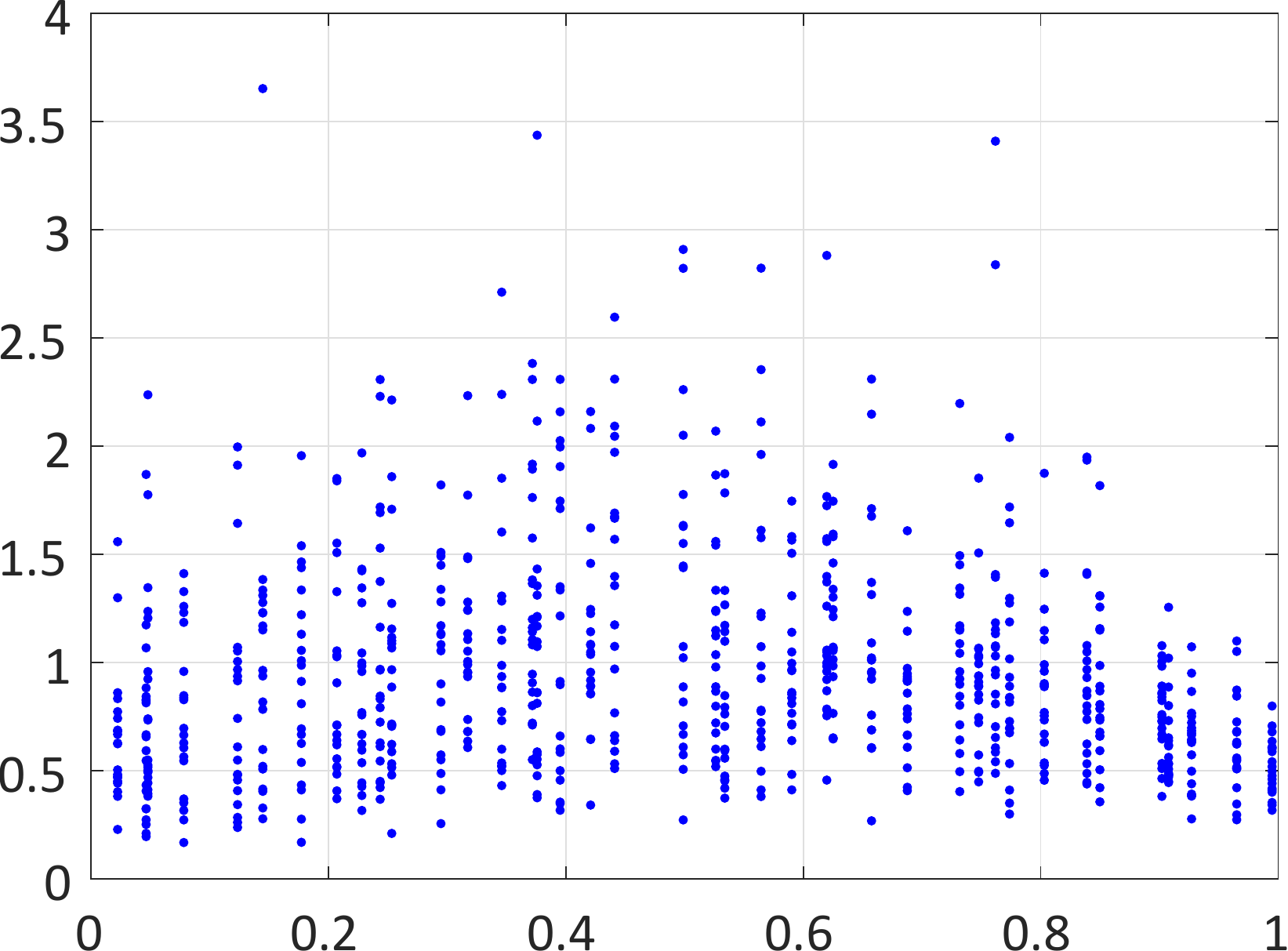}}
	\hspace{1cm}
	\subfigure[PDF prediction for $X=0.5$]{\includegraphics[height=.33\linewidth, keepaspectratio]{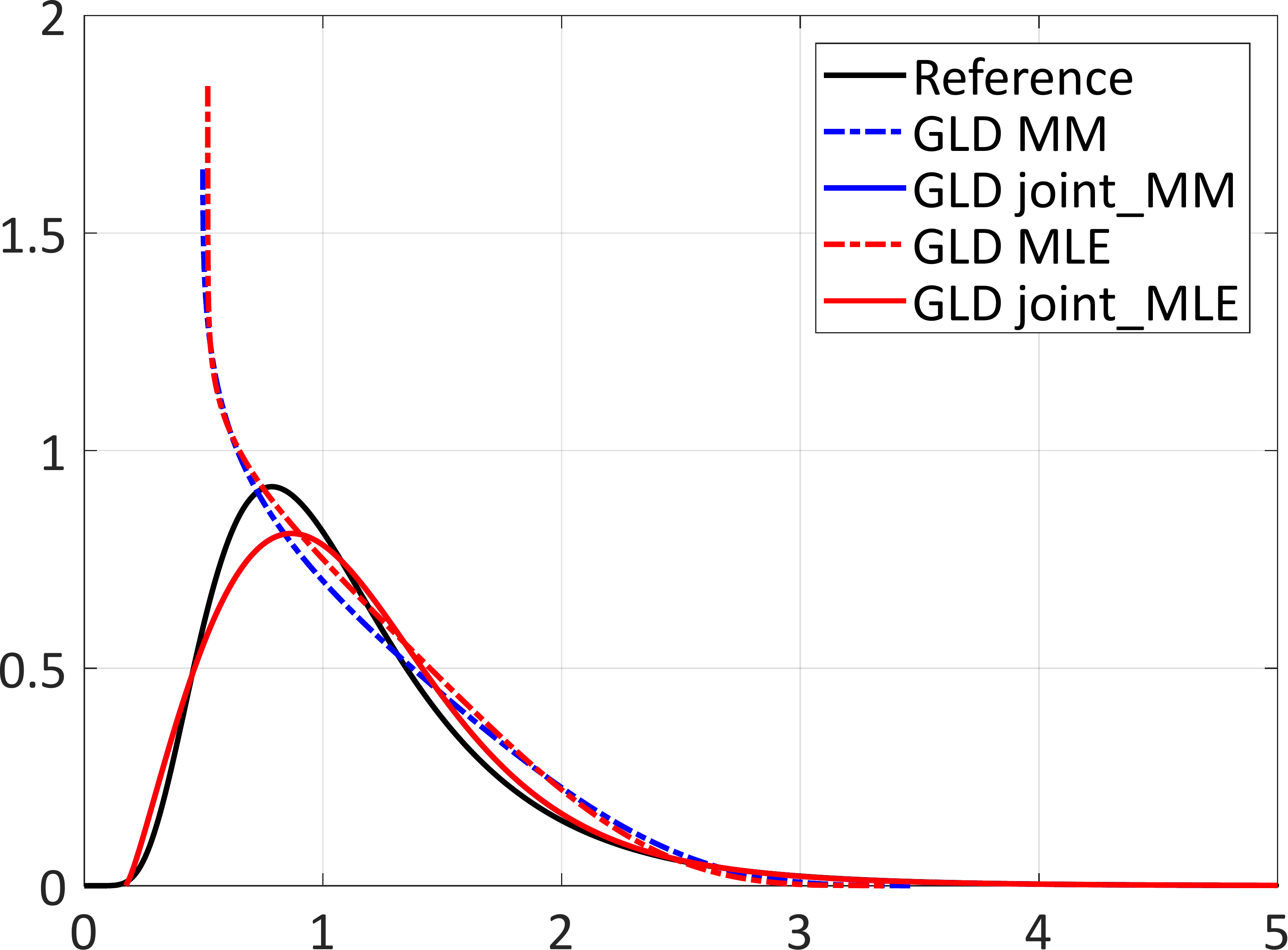}}
	\caption{Example 1 -- 40 ED points and 20 replications}
	\label{fig:logn_data}
\end{figure}

\Cref{fig:logn_data} shows one realization of an experimental design of $N=40$ and $R = 20$ for each point and the predicted PDF of the four surrogate models for $X=0.5$. We observe that the two models \GMM\ and \GMLE\ built using the Infer-and-Fit algorithm cannot capture the shape of the true distribution. In contrast, the two joint models \GJMM\ and \GJMLE\ produce a PDF that not only has the correct shape but also is an accurate approximation of the underlying distribution. We remark that with the data illustrated in \Cref{fig:logn_data}, \GJMM\ and \GJMLE\ are identical, implying that even though \GMM\ and \GMLE\ are different, their associated joint models can still be identical if they select the same sparse truncation sets $\acc{\caa_s, s=1,\ldots,4}$. Therefore, the selected algorithm is appears to be not too sensitive to the starting point.
\par
\par
\begin{figure}[ht!]
	\centering
	\includegraphics[width=.95\linewidth, keepaspectratio]{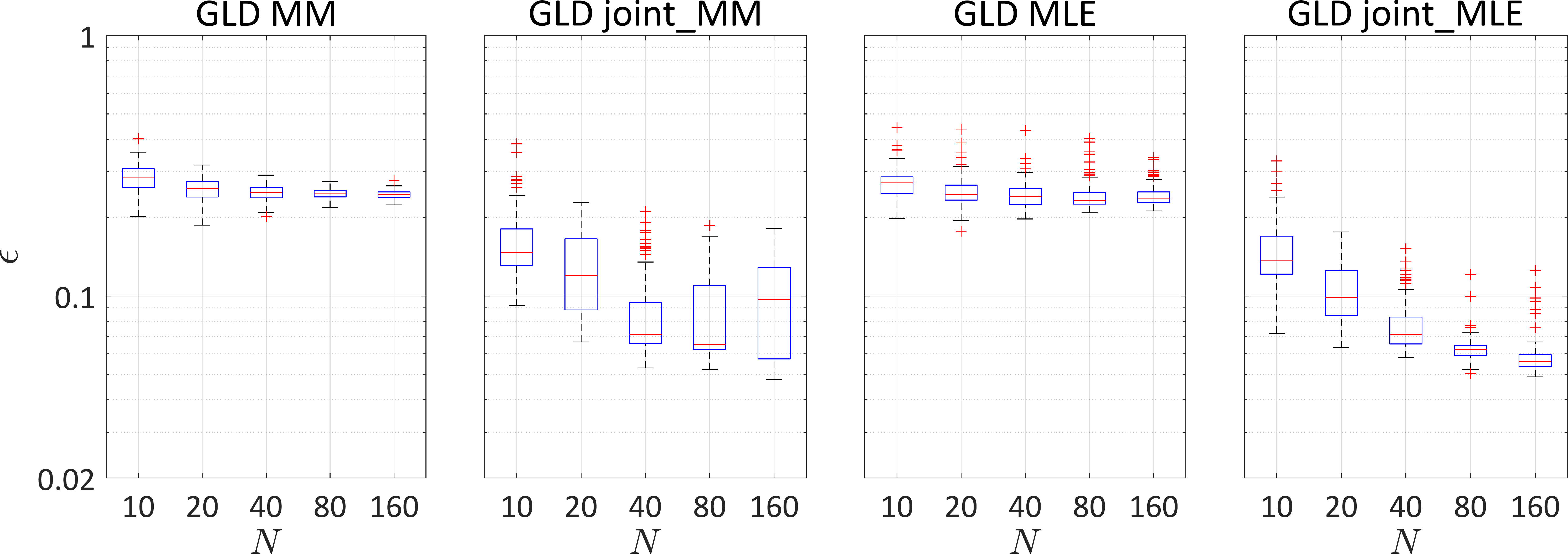}
	\caption{Example 1 -- Hellinger distance between the surrogate model built with $R = 20$ and the true response PDF, averaged over $\cx_{\rm test}$ (log-scale).}
	\label{fig:logn_result2}
\end{figure}
\begin{figure}[ht!]
	\centering
	\includegraphics[width=.95\linewidth, keepaspectratio]{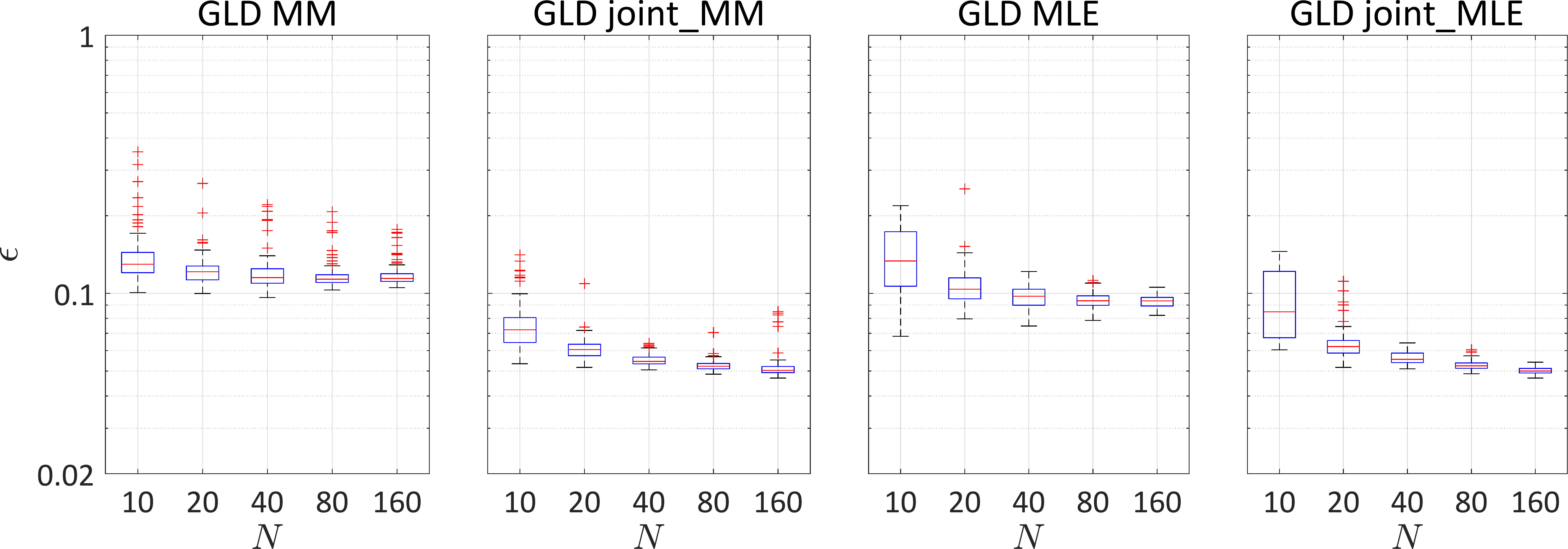}
	\caption{Example 1 -- Hellinger distance between the surrogate model built with $R = 80$ and the true response PDF, averaged over $\cx_{\rm test}$ (log-scale).}
	\label{fig:logn_result4}
\end{figure}

\Crefrange{fig:logn_result2}{fig:logn_result4} show quantitative comparisons of the convergence behavior of the four models. It turns out that in general all the GLD models converge when increasing the size of experimental design $N$ and the number of replications $R$. Moreover, the two joint models outperform the models built of the Infer-and-Fit approach, especially when only a few replications are available. 
\par
In the case with only 20 replications (\Cref{fig:logn_result2}), the convergence behavior of \GMM\ and \GMLE\ shows a weak dependence on $N$. This is because in the first step of \Cref{alg:twostep}, estimators $\hat{\ve{\lambda}}^{(i)}$ from both the method of moments and the maximum likelihood estimation might be biased. Then regression used in the second step is not able to filter the bias. Moreover, a few replications lead to high variance of the estimators, which together with the bias explains the non convergent behavior of \GMM\ and \GMLE. When increasing the number of replications, the bias becomes less significant and the variance of the error decreases. Therefore, $\epsilon$ decreases with increasing $N$ for \GMM\ and \GMLE\ in \Cref{fig:logn_result4}. 
\par
In contrast, \GJMLE\ exhibits a fast error decay even with a small number of replications. This is because all the available data are used at once to estimate the model parameters, which reduces both the bias and the variance. \GJMM\ appears to provide less accurate PDF estimation than \GJMLE, which is due to the less appropriate truncation scheme selected by \GMM. Nevertheless, \GJMM\ still improves the results of \GMM\ and outperforms \GMLE. 
\par
\par
\begin{figure}[ht!]
	\centering
	\includegraphics[width=.95\linewidth, keepaspectratio]{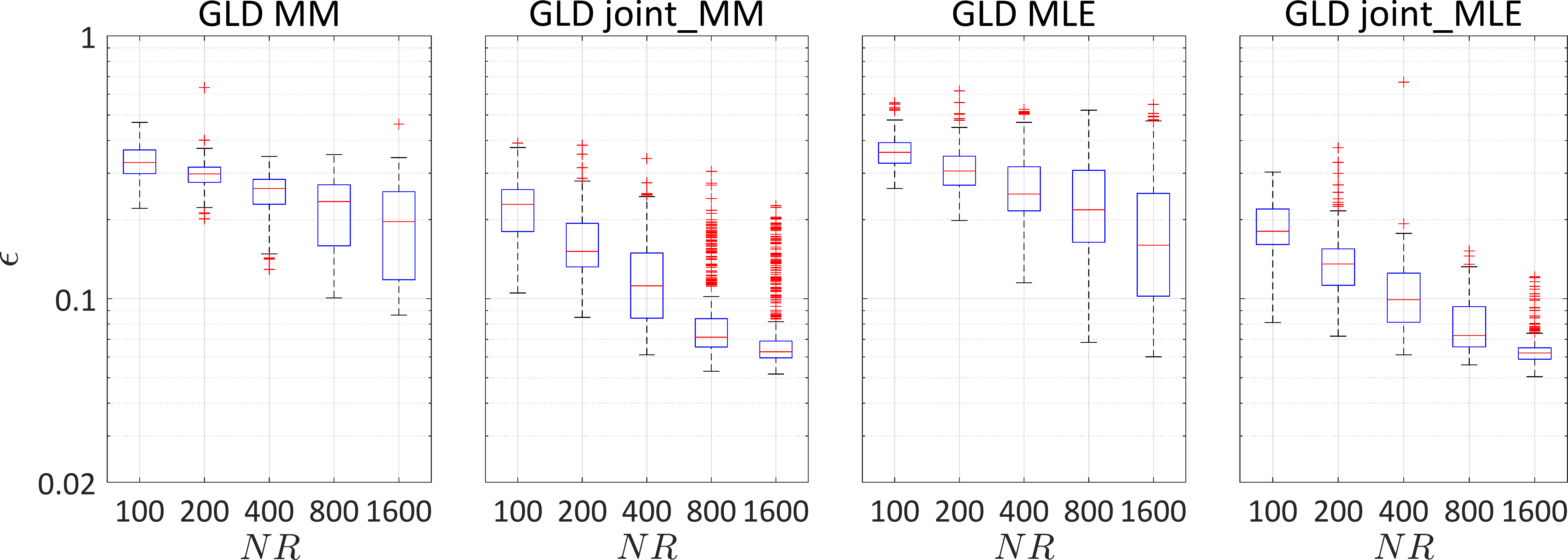}
	\caption{Example 1 -- Hellinger distance between surrogate models built with different total number of model runs and the true response PDF, averaged over $\cx_{\rm test}$ (log-scale).}
	\label{fig:logn_resultot}
\end{figure}

We have run the simulation for $N = \acc{10,20,40,80,160}$ and $R = \acc{10,20,40,80,160}$. \Cref{fig:logn_resultot} summarizes the total number of model runs $NR$ (only up to 1,600) against the error measure. The results are consistent with what we have observed in the case of fixed number of replications. More precisely, the two models built with the joint modeling algorithm outperform those based on the Infer-and-Fit algorithm: with 400 models runs, \GJMM\ and \GJMLE\ provide more accurate PDF estimations than \GMM\ and \GMLE\ with 1,600 model evaluations.
\par

\subsection{Example 2: a five-dimensional simulator}

The second analytical example is defined as follows:
\begin{align}
Y\left(\ve{X},\omega\right) = \mu(\ve{X})+\sigma(\ve{X}) \cdot Z(\omega),
\end{align}
where $Z(\omega) \sim \mathcal{N}(0,1)$ is the latent variable that represents the source of randomness, and $\ve{X}$ is a five-dimensional random vector, with independent components having uniform distribution $\cu(0,1)$. $Y(\ve{x})$ is a Gaussian random variable with mean $\mu(\ve{x})$ and standard deviation $\sigma(\ve{x})$. In this example, the mean function $\mu(\ve{x})$ reads
\begin{align}
\mu(\ve{x}) = 3-\sum_{j=1}^{5}jx_j + \frac{1}{5}\sum_{j=1}^{5}jx^3_j +  \frac{1}{15}\log\left(1+\sum_{j=1}^{5}j(x^2_j+x^4_j)\right) + x_1 \, x^2_2 - x_5 \, x_3 + x_2 \, x_4,
\end{align}
and the standard deviation $\sigma(\ve{x})$ is given by
\begin{align}
\sigma(\ve{x}) = \exp\left(\frac{1}{4}\sum_{j=1}^{5}x_j\right),
\end{align}
which implies a strong heteroskedastic effect with a highly nonlinear mean function. This example is used to show the performance of the proposed methods in moderate dimensional problems.
\par
\begin{figure}[!htbp]
	\centering
	\subfigure[PDF at $\ve{x} = \left(0.25,0.25,0.25,0.25,0.25\right)^T$]{\includegraphics[height=.33\linewidth, keepaspectratio]{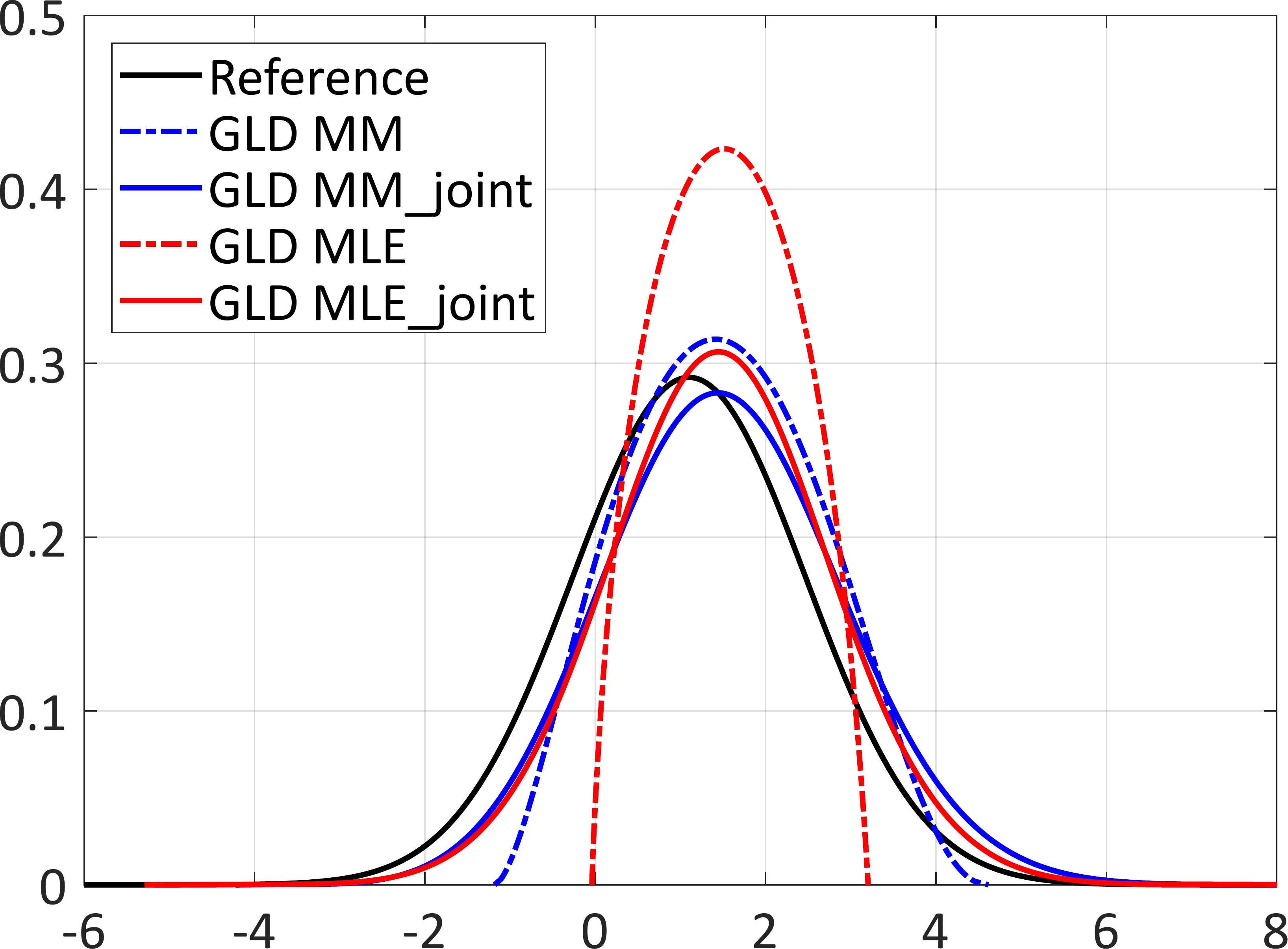}}
	\hspace{1cm}
	\subfigure[PDF at $\ve{x} = \left(0.75,0.75,0.75,0.75,0.75\right)^T$]{\includegraphics[height=.33\linewidth, keepaspectratio]{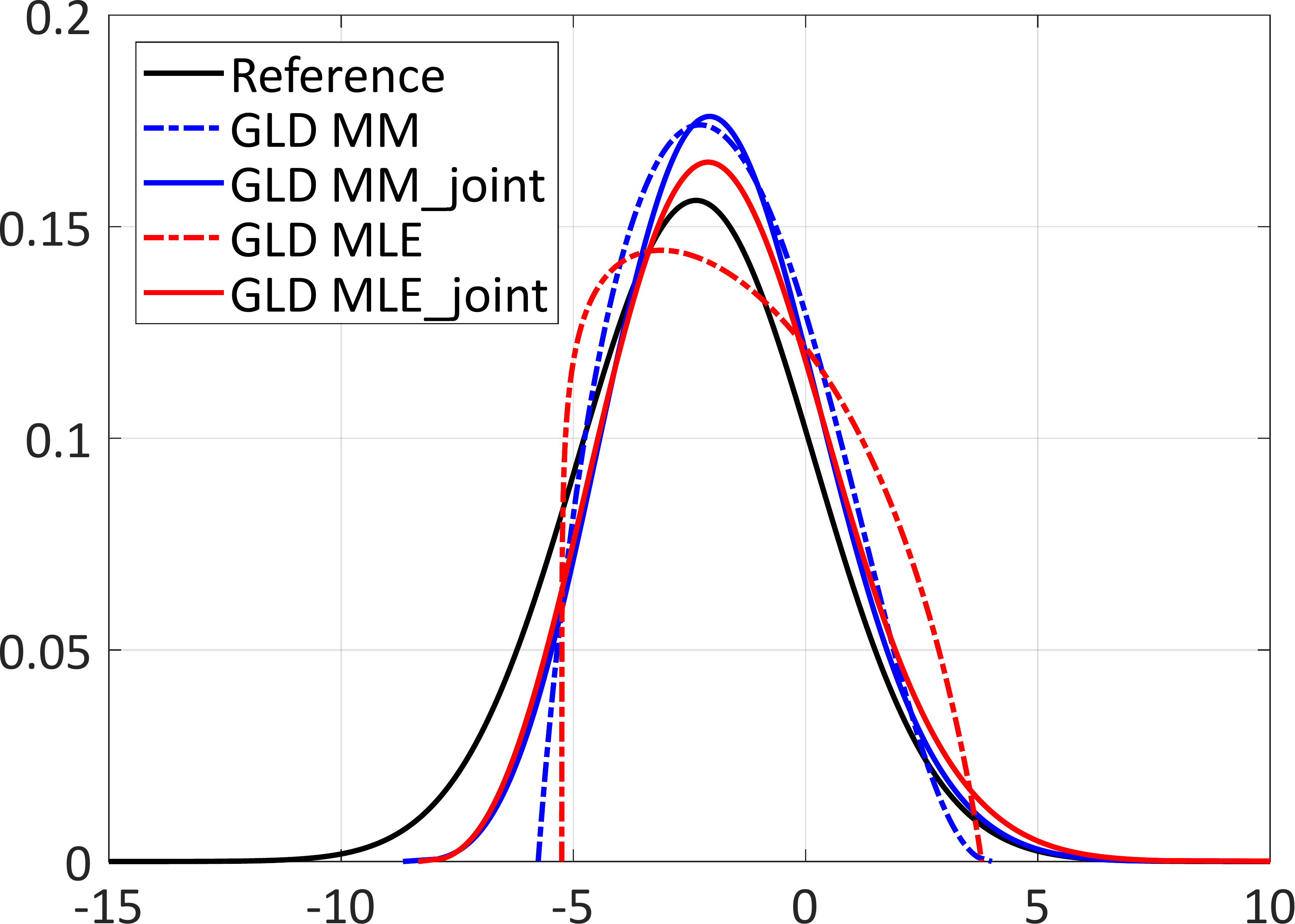}}
	\caption{Example 2 -- PDF predictions with an experimental design of size 50 and 25 replications.}
	\label{fig:normpdf}
\end{figure}
\par
\begin{figure}[ht!]
	\centering
	\includegraphics[width=.95\linewidth, keepaspectratio]{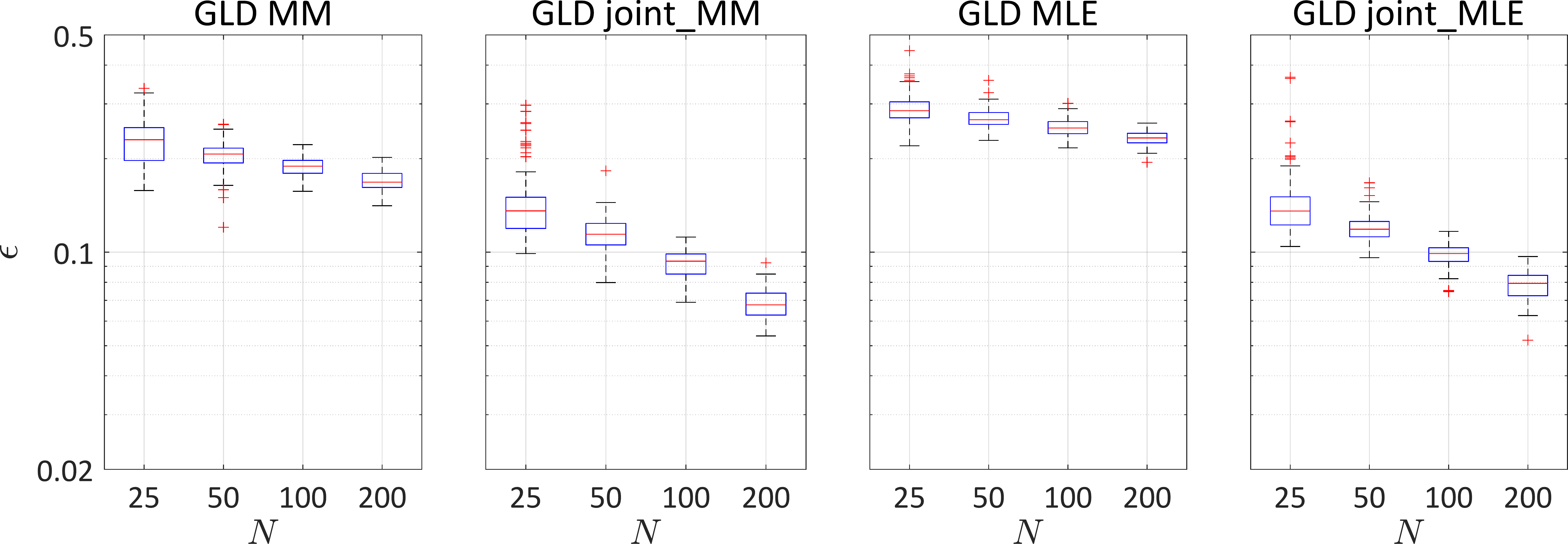}
	\caption{Example 2 -- Hellinger distance between the surrogate model built with $R = 25$ and the true response PDF, averaged over $\cx_{\rm test}$ (log-scale)}
	\label{fig:norm_result1}
\end{figure}
\begin{figure}[ht!]
	\centering
	\includegraphics[width=.95\linewidth, keepaspectratio]{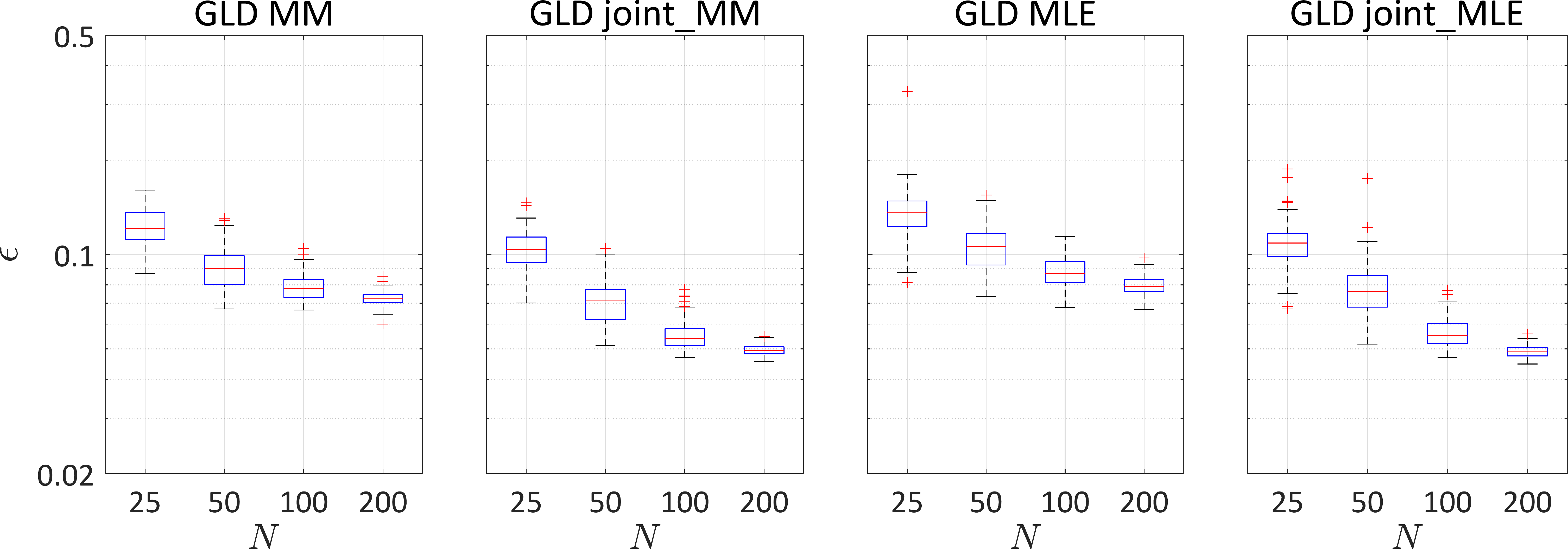}
	\caption{Example 2 -- Hellinger distance between the surrogate model built with $R = 100$ and the true response PDF, averaged over $\cx_{\rm test}$ (log-scale)}
	\label{fig:norm_result3}
\end{figure}
\begin{figure}[ht!]
	\centering
	\includegraphics[width=.95\linewidth, keepaspectratio]{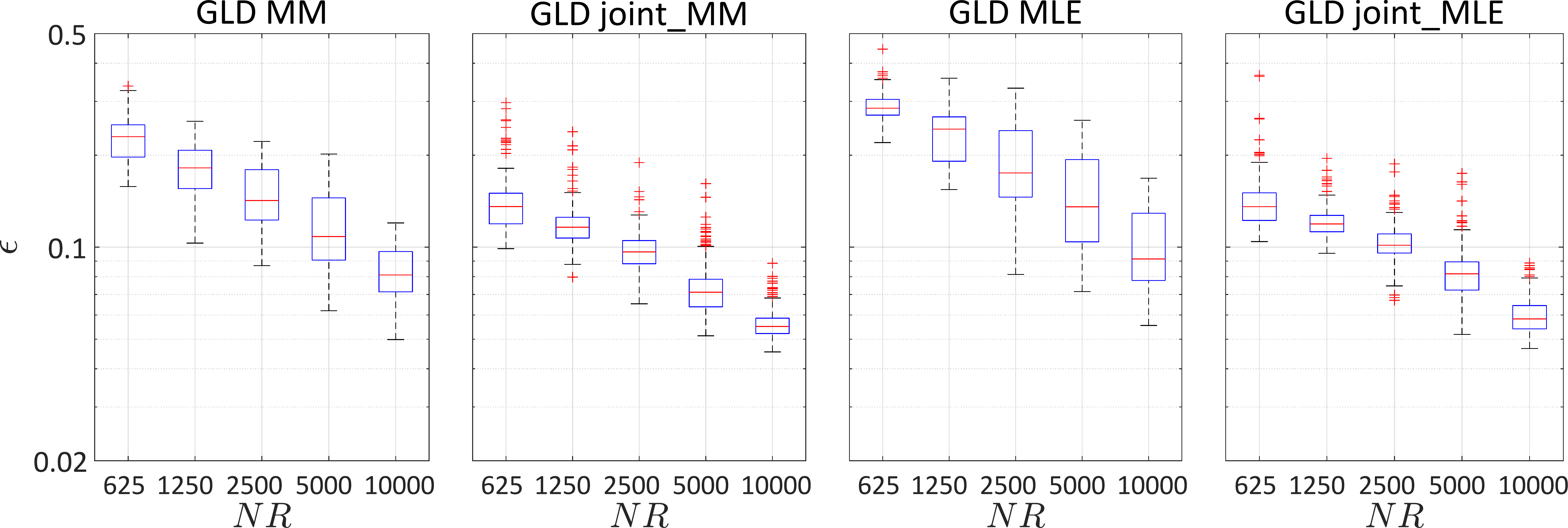}
	\caption{Example 2 -- Hellinger distance between surrogate models built with different total number of model runs and the true response PDF, averaged over $\cx_{\rm test}$ (log-scale).}
	\label{fig:norm_resultot}
\end{figure}

Similar to the previous example, the GLD models demonstrate a convergent behavior, as illustrated in \Crefrange{fig:norm_result1}{fig:norm_resultot}. The two joint models yield more accurate estimates than those built with the Infer-and-Fit algorithm. In the case of a few replications, both \GMM\ and \GMLE\ fail to capture the shape of the PDF (\Cref{fig:normpdf}), and thus converge rather slowly with respect to $N$ (see \Cref{fig:norm_result1}). In contrast, the two joint models are less sensitive to the number of replications, and their performance mainly depends on the ED size. Unlike the first example, using the method of moment turns out to produce slightly more accurate estimates when applying the Infer-and-Fit algorithm. However, the parametric estimation methods employed to get $\hat{\ve{\lambda}}^{(i)}$ do not have a significant influence on the accuracy of the joint algorithm: \GJMM\ and \GJMLE\ show very similar convergence. 
\par
\par
In this section, only the error measure based on the Hellinger distance is reported for convergence studies. Nevertheless, quantitative comparisons using other metrics such as the Kolmogorov-Smirnov distance, the mean value and the 95\% quantile of the predicted distributions show similar trends.

\section{Applications}
\label{sec:appli}
\subsection{Stochastic differential equation}
\label{sec:sde}
Stochastic differential equations (SDE) are widely used to model the evolution of complex systems in many fields, \eg finance \citep{McNeil2005}, epidemics \citep{Gray2011}, and meteorology \citep{Iversen2015}. Due to the stochastic process (\eg Wiener processes) involved in a SDE, the associated solution is also a stochastic process. As a results, when fixing the parameters of a SDE, any scalar-valued deterministic function of the solution process produces a random variable, which can be regarded as a stochastic simulator. In this case study, we consider the example proposed by \cite{Jimenez2017}, the governing equation of which reads
\begin{align}
\D Y_t = (X_1 - Y_t)\D t + (\nu Y_t + 1)X_2 \, \D W_t, \label{eq:defsde}
\end{align}
with the initial condition defined by
\begin{align}
Y_0 = 0 \,\,\, \text{almost surely}. \nonumber
\end{align}
In this equation, $\ve{X} = \left(X_1,X_2\right)^T$ are the SDE parameters, and $W_t$ is a standard Wiener process that represents the source of randomness. We denote $Y_t(\ve{x})$ the solution of \refEq{eq:defsde} for $\ve{X} = \ve{x}$, and we focus on the value of $Y_t(\ve{x})$ at $t = 10$, \ie $Y_{10}(\ve{x})$ is the scalar QoI.
\par
Note that the value of $\nu$ controls how the Wiener process affects the QoI: for $\nu = 0$, $\D W_t$ is multiplied with a constant, and thus the solution $Y_t(\ve{x})$ is a Gaussian process; whereas for $\nu \neq 0$, $W_t$ interacts with the unknown process $Y_t(\ve{x})$, and the marginal distribution of $Y_t(\ve{x})$ does not have an analytical closed-form. We set $\nu = 0.2$ in this study. To numerically solve \refEq{eq:defsde}, we apply the classical Euler-Maruyana method \citep{Kloeden1992} with time step $\Delta t= 0.01$. Therefore, the discretized version of \refEq{eq:defsde} has a large number of latent random variables $\ve{Z}$ equal to $\frac{10}{\Delta t} = $1,000. This problem is representative of cases with low dimensionality in $\ve{X}$ and very large size of $\ve{Z}$.
\par
The original definition of $\ve{X}$ proposed by \cite{Jimenez2017} follows $X_1 \sim \cu(0.95,1.15)$ and $X_2 \sim \cu(0.02,0.22)$. According to some preliminary tests, we found that under this setting, the response PDF is close to a normal distribution and does not vary significantly with respect to $\ve{x}$ because the range of definition of the input parameters is rather narrow. In order to have richer shapes for the output PDF of $Y_{10}(\ve{x})$ and challenge our algorithm, we choose $X_1 \sim \cu(0.9,2)$, $X_2 \sim \cu(0.1,1)$ in this paper. Thus, the response PDF can have normal-like shape and can also be right-skewed depending on $\ve{x}$. 
\par
\begin{figure}[!htbp]
	\centering
	\subfigure[PDF at $X = \left(1.2,0.3\right)^T$]{\includegraphics[height=.33\linewidth, keepaspectratio]{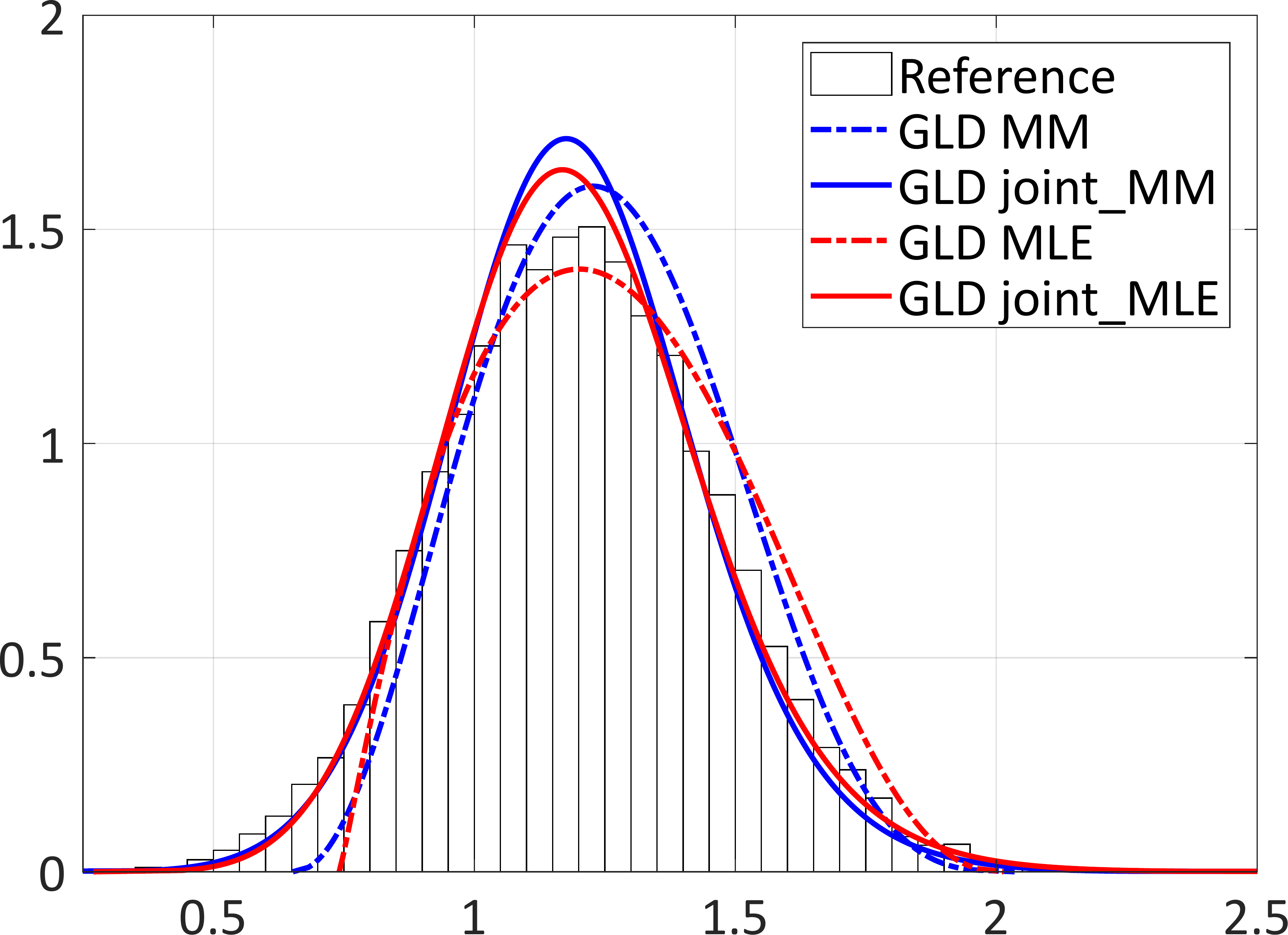}}
	\hspace{1cm}
	\subfigure[PDF at $X = \left(1.8,0.8\right)^T$]{\includegraphics[height=.33\linewidth, keepaspectratio]{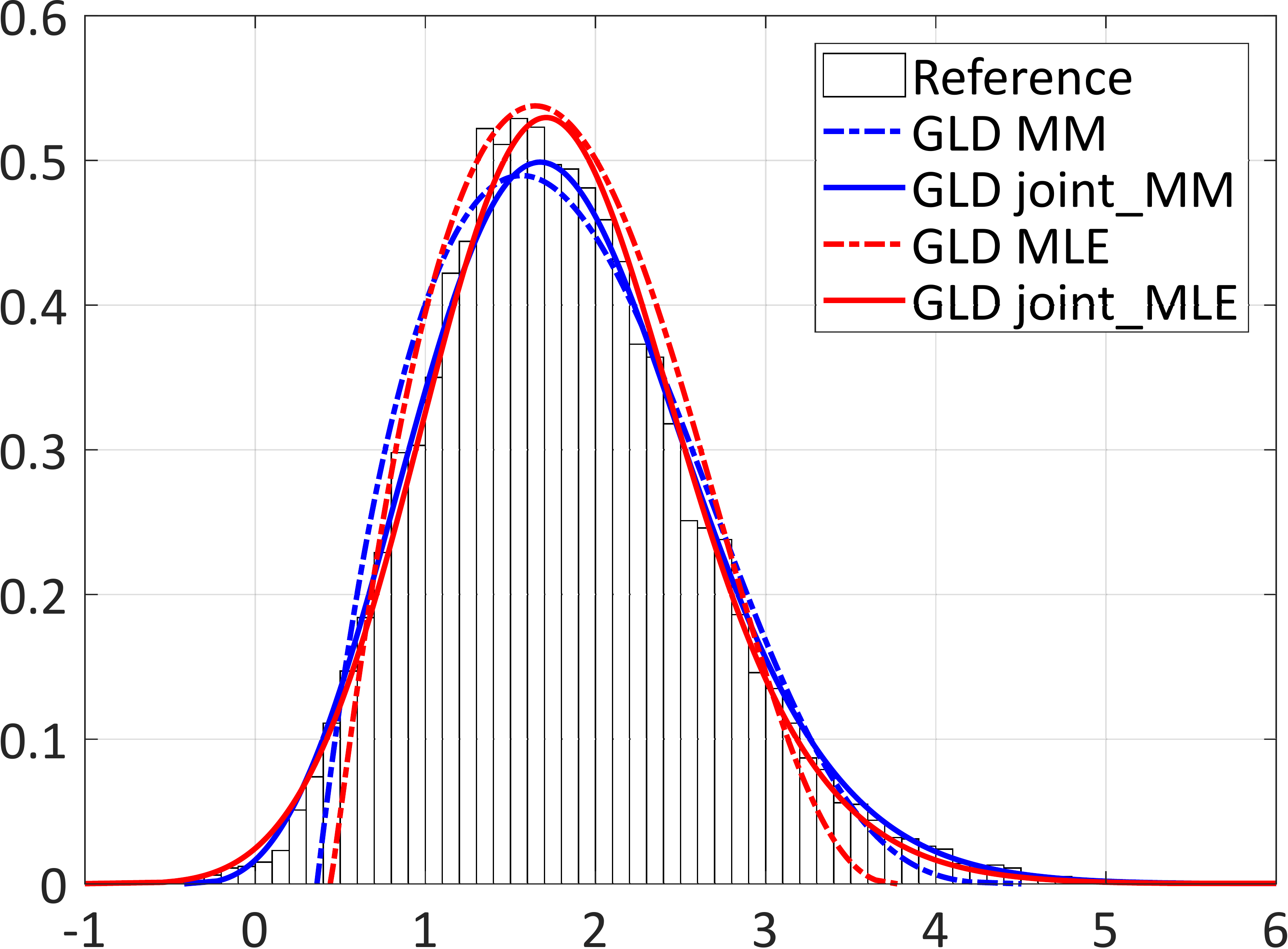}}
	\caption{Stochastic differential equation -- PDF predictions with an experimental design of size 80 using 40 replications. The reference histogram is calculated based on 10,000 replications.}
	\label{fig:SDEpdf}
\end{figure}
\par
\Cref{fig:SDEpdf} shows the results when applying the developed methods to an experimental design of $N=80$ and $R = 40$. We observe that all the four models can generally well approximate the underlying distributions. Detailed comparison shows that the Infer-and-Fit algorithm is not able to correctly emulate the shape variation of the response PDF: when the underlying PDF is close to a normal distribution, both \GMM\ and \GMLE\ predict a slightly right-skewed PDF; whereas for positively skewed PDF, neither of them is able to accurately approximate the tail. In contrast, \GJMM\ and \GJMLE\ not only capture the shape variation but also better represent the underlying PDF.
\par
\begin{figure}[ht!]
	\centering
	\includegraphics[width=.95\linewidth, keepaspectratio]{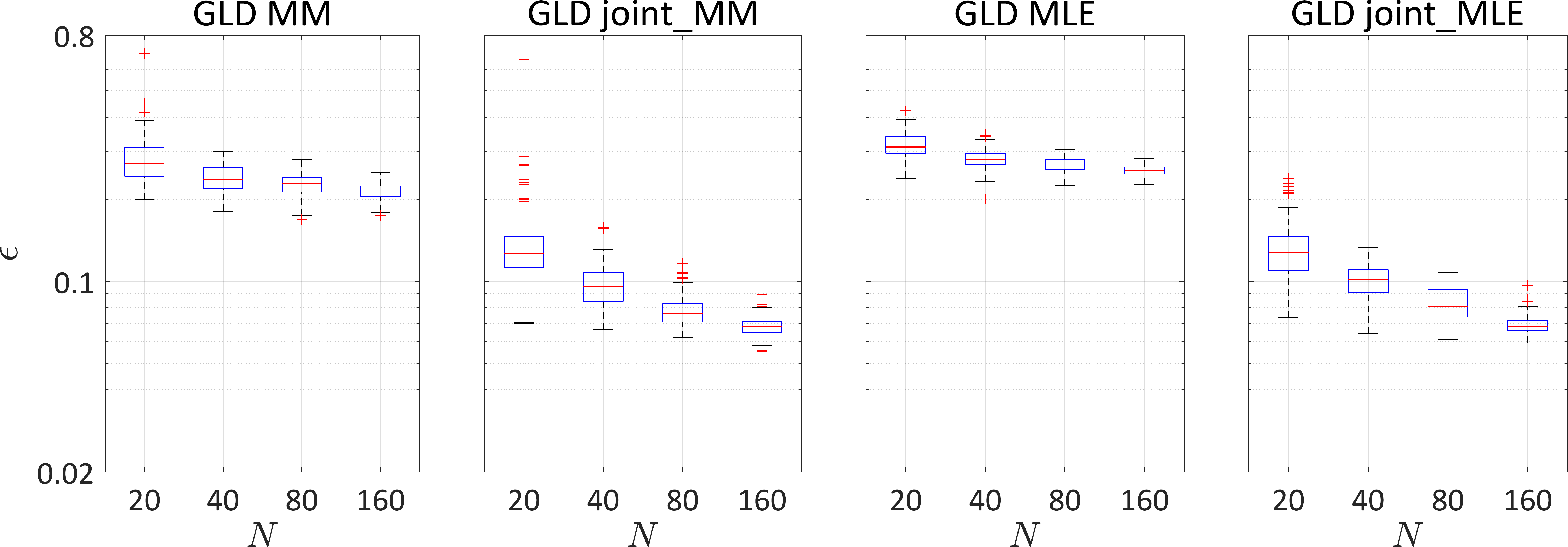}
	\caption{Stochastic differential equation -- Hellinger distance between the surrogate model built with $R = 20$ and the reference response PDF, averaged over $\cx_{\rm test}$ (log-scale).}
	\label{fig:SDE_result1}
\end{figure}
\begin{figure}[ht!]
	\centering
	\includegraphics[width=.95\linewidth, keepaspectratio]{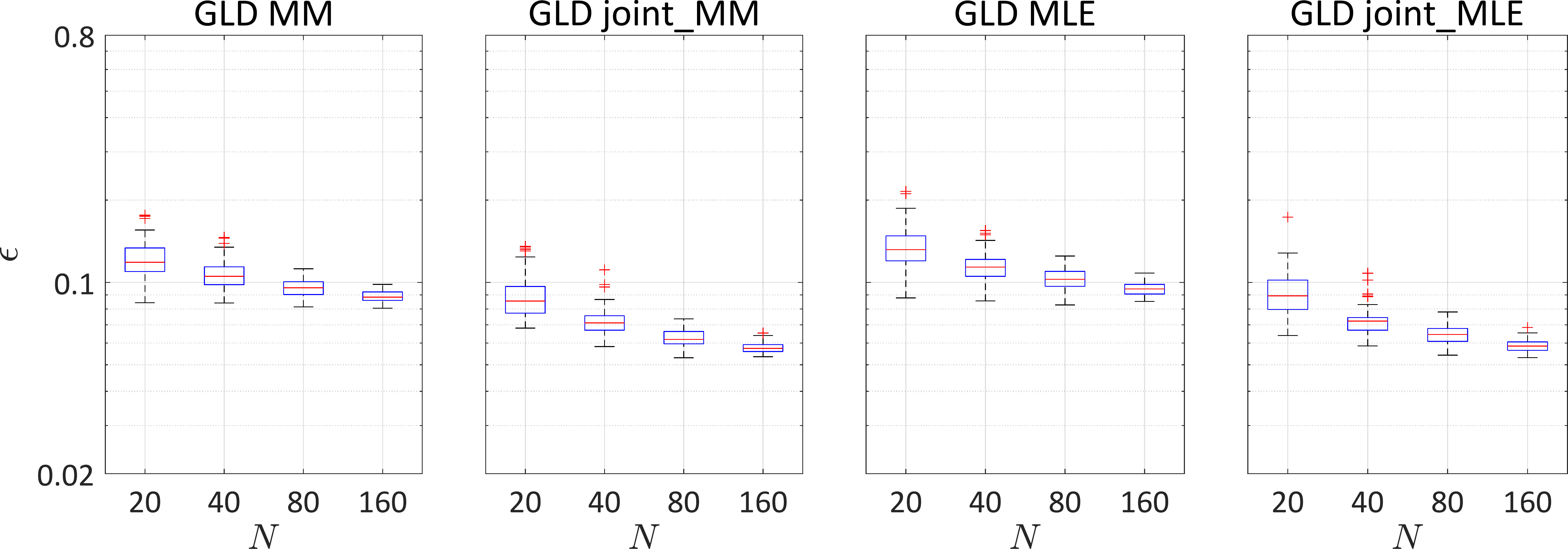}
	\caption{Stochastic differential equation -- Hellinger distance between the surrogate model built with $R = 80$ and the reference response PDF, averaged over $\cx_{\rm test}$ (log-scale).}
	\label{fig:SDE_result3}
\end{figure}
\par
Similar to the analytical examples in \Cref{sec:examples}, we investigate the convergence behavior of the developed methods. Since the analytical PDF is not available, we use kernel density estimation \citep{Wand1995} using 10,000 replications as the reference distribution for each point in the test set. The Hellinger distance between the predicted PDF and the reference is averaged over a test set $\cx_{\rm test}$ containing 100 points generated with a Sobol' sequence. 
\par
The convergence study of the four models is reported in \Crefrange{fig:SDE_result1}{fig:SDE_result3}. As expected from the analytical examples, the joint modeling algorithm appears much more efficient. In particular, both \GJMM\ and \GJMLE\ yield an error around 0.07 in the case of 20 replications and 80 ED points, \ie 1,600 model evaluations, whereas \GMM\ and \GMLE\ can barely achieve this accuracy even when 80 replications and 160 ED points, \ie a total of 12,800 model evaluations, are available. More generally, the joint algorithm produce more accurate results than the Infer-and-Fit algorithm when a large number of replications are available. 

\subsection{Wind turbine design}
\label{sec:wind}
\begin{figure}[!b]
	\centering
	\includegraphics[width=0.95\linewidth]{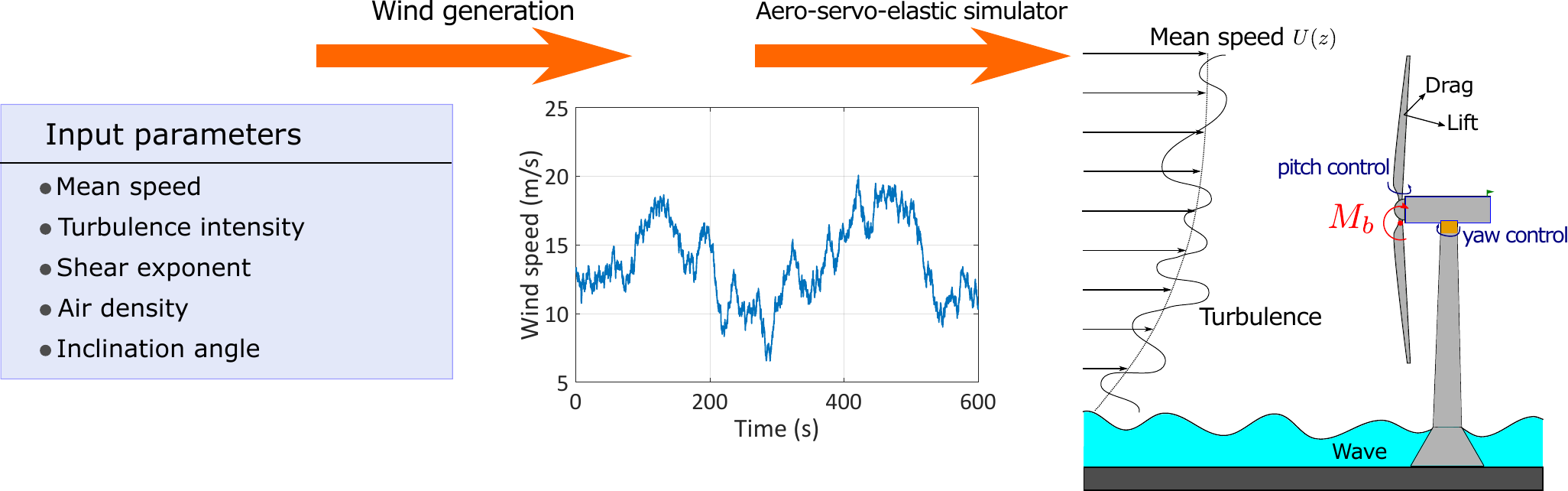}
	\caption{Wind turbine simulation scheme}
	\label{fig:simulator1}
\end{figure}

In the wind turbine design process, structural components need to be analyzed under diverse environmental loads to assess their performance and reliability. Typical simulations consist of two parts, namely the generation of the external excitations (\ie wind inflow) and the aero-servo-elastic simulation as illustrated in \Cref{fig:simulator1}. The latter refers to the complex multi-physics scenario including mutual interactions of wind inflow, aerodynamics, structural dynamics (elastic deflections) and control systems.
\par
\par
The wind field generator used in this study is TurbSim \citep{nrel2009turbsim}, which is a stochastic inflow turbulence simulator. It takes five macroscopic parameters as input: (1) the mean wind velocity $U$ at reference altitude $z_{\rm ref}$; (2) the turbulence intensity $I$, denoting the coefficient of variation of the wind time series, \ie $I = \sigma/U$; (3) the wind shear exponent $\alpha$, describing the variation of the mean wind velocity with the altitude according to the following equation:
\begin{equation}
U(z) = U\cdot\left(\frac{z}{z_{\rm ref}}\right)^{\alpha};
\end{equation} 
(4) the air density $\rho$ and (5) the inclination angle $\beta$ (see \cite{AbdallahPEM2019} for details). Since these five parameters cannot fully determine a wind field, random seeds are used on top of these macroscopic parameters in TurbSim to generate a coherent turbulent three-dimensional velocity time series \citep{nrel2009turbsim}. 
\par
The wind turbine structure studied here is the reference 5~MW upwind turbine described in \cite{nrel2009definition}. The aero-servo-elastic simulator is FAST \citep{nrel2009definition}, a deterministic computational model that takes inflow wind fields as input and calculates the structural response as output. However, due to the use of random seeds in the turbulent wind generation, simulations of wind turbines are stochastic with respect to the five input macroscopic parameters. In other words, fixing the five quantities described above, any number of three-dimensional wind fields can be simulated, each of which leads to a different response and predicted performance of the wind turbine. Note that this is also what happens in reality for wind turbines.
\par
\begin{figure}[!htbp]
	\centering
	\subfigure[Bounds in $(I,U)$ plane]{\includegraphics[height=.253\linewidth, keepaspectratio]{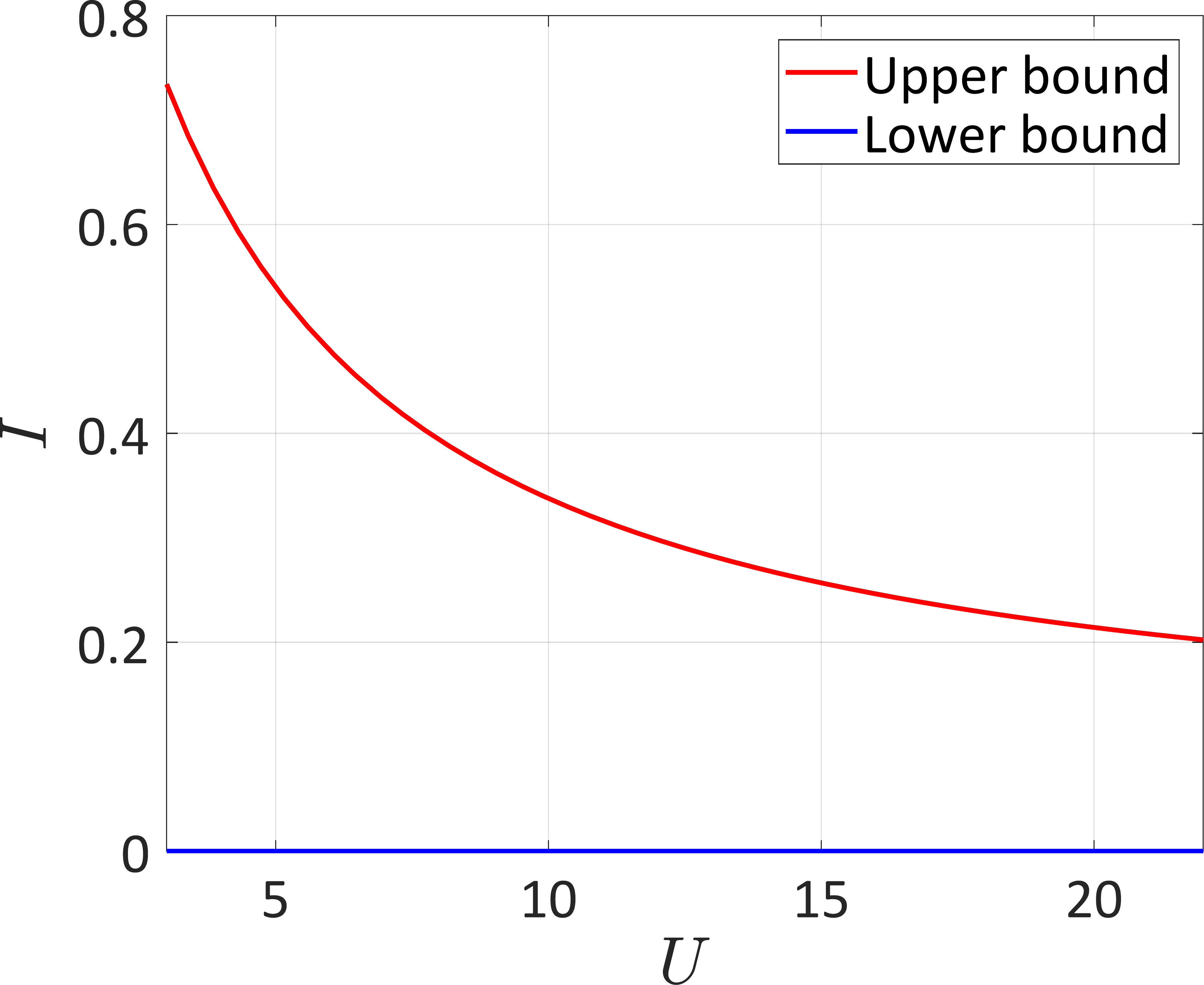}}
	\hspace{0.3cm}
	\subfigure[Bounds in $(\alpha,U)$ plane]{\includegraphics[height=.25\linewidth, keepaspectratio]{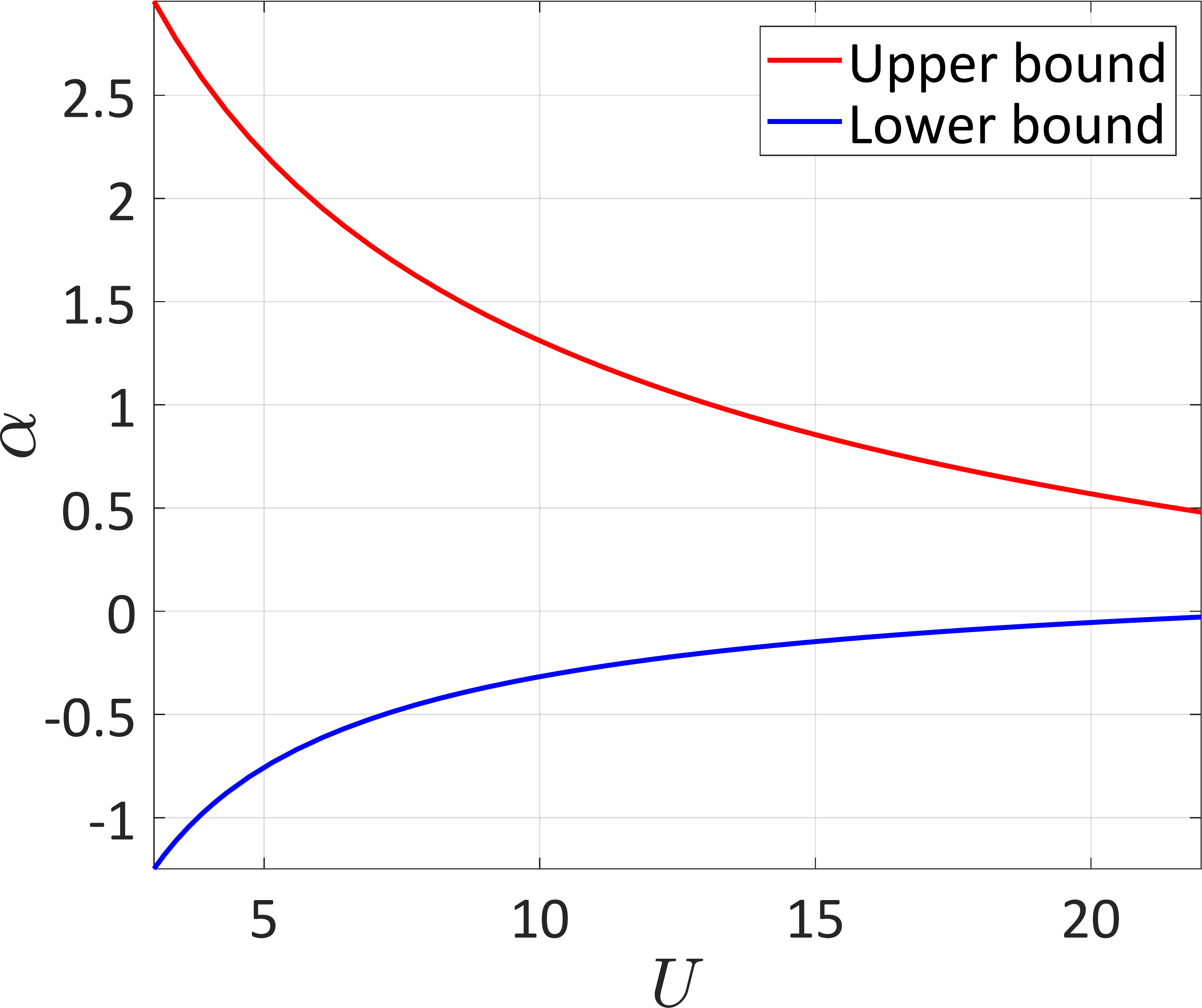}}
	\hspace{0.3cm}
	\subfigure[Bounds in $(\alpha,I)$ plane]{\includegraphics[height=.25\linewidth, keepaspectratio]{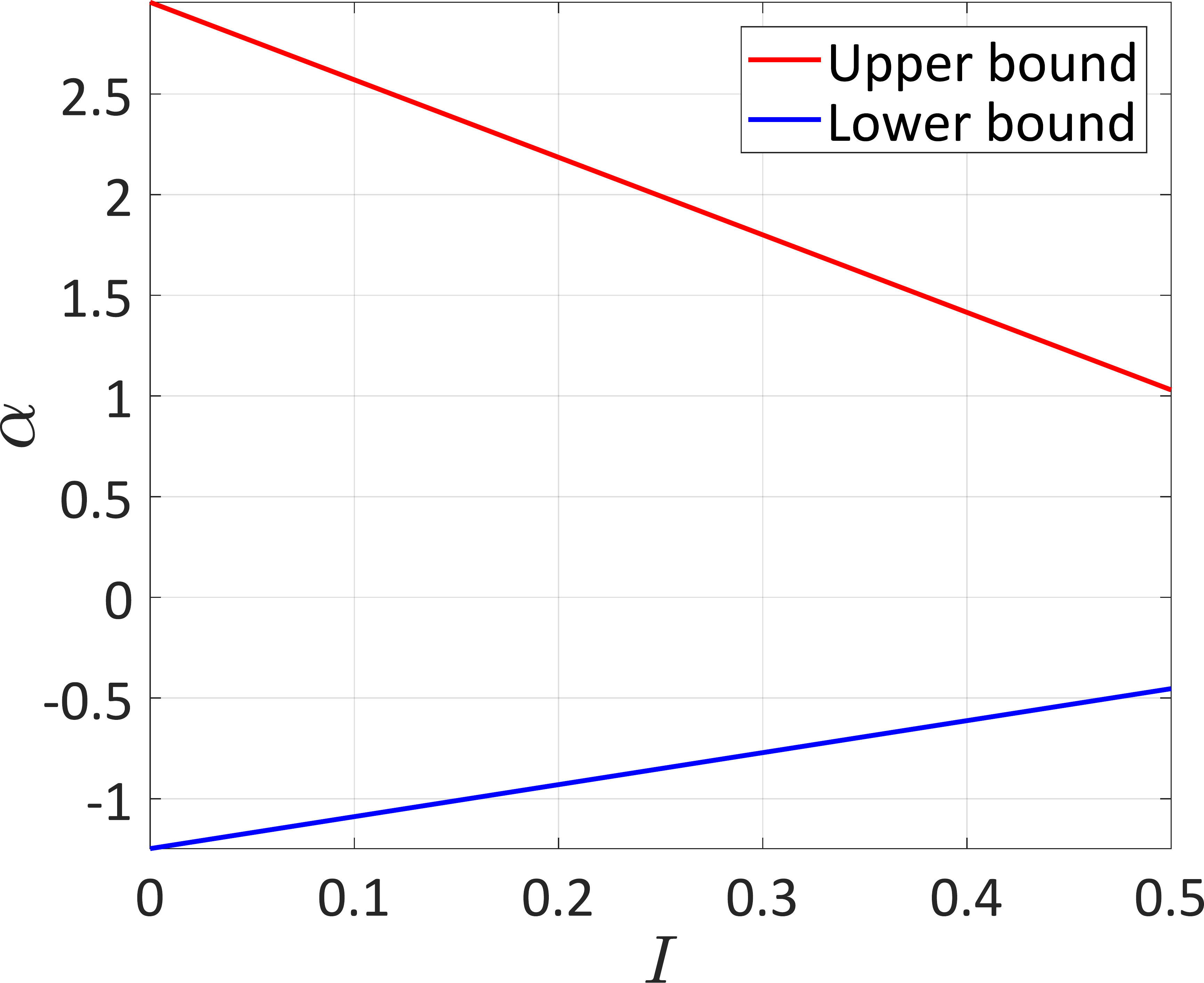}}
	\caption{Bounds on the physical parameters $(U,I,\alpha)$ calibrated from real wind data}
	\label{fig:windbounds}
\end{figure}

Of interest is the maximum flap-wise bending moment at the blade root $M_b$ within the simulated time (10 minutes) for a given wind climate defined by the 5 macroscopic parameters, as illustrated in \Cref{fig:simulator1}. To build a stochastic surrogate, the Latin hypercube sampling (LHS) method \citep{McKay1979} with rejection is used to create an experimental design of 485 points in dimension 5. More precisely, input samples are firstly generated by the LHS following \Cref{tab:windinput}, and then the samples that are outside the bounds that are calibrated from real wind climate are removed. The bounds are respectively defined in the $(I,U)$ plane, $(\alpha,U)$ plane, and $(\alpha,I)$ plane, as illustrated in \Cref{fig:windbounds} \citep{SlotWind2020}. Importantly, when the turbulence standard deviation $\sigma= I\cdot U$ is close to zero, the wind speed barely varies in time, and thus the response PDF is close to being degenerate, which can cause numerical problems. In this case, the simulator can be considered as deterministic and does not fit to the GLD framework. Hence, we introduce an additional bound in $\sigma$, and only samples with $\sigma>0.05$ are simulated. Finally, the simulator is run 50 times for each set of input parameters as replications. 
\par
Considering the physical process, we chose to use the turbulence standard deviation $\sigma$ rather than the turbulence intensity $I$ to build the surrogate models. Hence, the input parameters are pre-processed as $\ve{X} = \left( U,\sigma, \alpha, \beta, \rho \right)^T$ for training. 
\par
\begin{table}
	\centering
	\caption{Wind turbine case study--description of the input variables}
	\begin{tabular}{  c c c c }
		\hline
		Name & Description & Distribution & Parameters \\
		\hline
		$U$ & Mean speed (m)& Uniform & $[3,22]$ \\
		$I$ & Turbulence intensity & Uniform & $[0,0.5]$ \\ 
		$\alpha$ & Shear exponent & Uniform & $[-2, 5]$ \\
		$\rho$ & Air density (kg$/$m$^3$) & Uniform & $[0.8,1.4]$ \\
		$\beta$ & Declination angle (deg) & Uniform & $[-10,10]$ \\
		\hline
	\end{tabular}
	\label{tab:windinput}
\end{table}
\par
Because of the sampling scheme, the input parameters $U$, $I$ and $\alpha$ are not independent, which violates the independent assumption when building PC basis. One possibility to tackle this problem would be to use the Rosenblatt transform \citep{Rosenblatt1952} to map the dependent inputs into a set of independent random variables and then build PC basis of the latter. However, \cite{Torre2019} shows that this approach, while yielding improved estimates of the output statistics, is typically detrimental to the accuracy of pointwise predictions. This is because the Rosenblatt transform is highly nonlinear, resulting in a transformed model whose PCE spectrum decays typically more slowly than the original one. Therefore, we ignore the dependence when building PC basis functions, and only the marginal distribution of each input variables is needed. Since the marginal distributions are difficult to be derived analytically due to the bounds, we apply the kernel density estimators to 10,000 samples generated according to the rejection sampling scheme described before. Note that the air density $\rho$ and the inclination angle $\beta$ are uniform variables, and they are independent from $U$, $I$ and $\alpha$. Therefore, we use Legendre polynomials as the associated univariate PC basis functions for these two variables.
\par
Unlike the previous example in \Cref{sec:sde}, the wind turbine simulation is costly, and thus we cannot run as many times the simulator as needed to have a reliable estimate of the error defined in \Cref{eq:Rlevel1}. To assess the performance of the proposed methods, 120 samples of input parameters are generated by the same scheme as the training set. The simulator is repeatedly run 500 times for each test point. We then compare some sample statistics with those predicted by the stochastic emulators built by the developed methods. The former are considered as references. The metrics used for comparison are the mean, the standard deviation (std) and the 5\%, 10\%, 50\%, 90\% and 95\% quantiles. 
\par
\begin{figure}[!htbp]
	\centering
	\subfigure[Mean estimation]{\includegraphics[height=.4\linewidth, keepaspectratio]{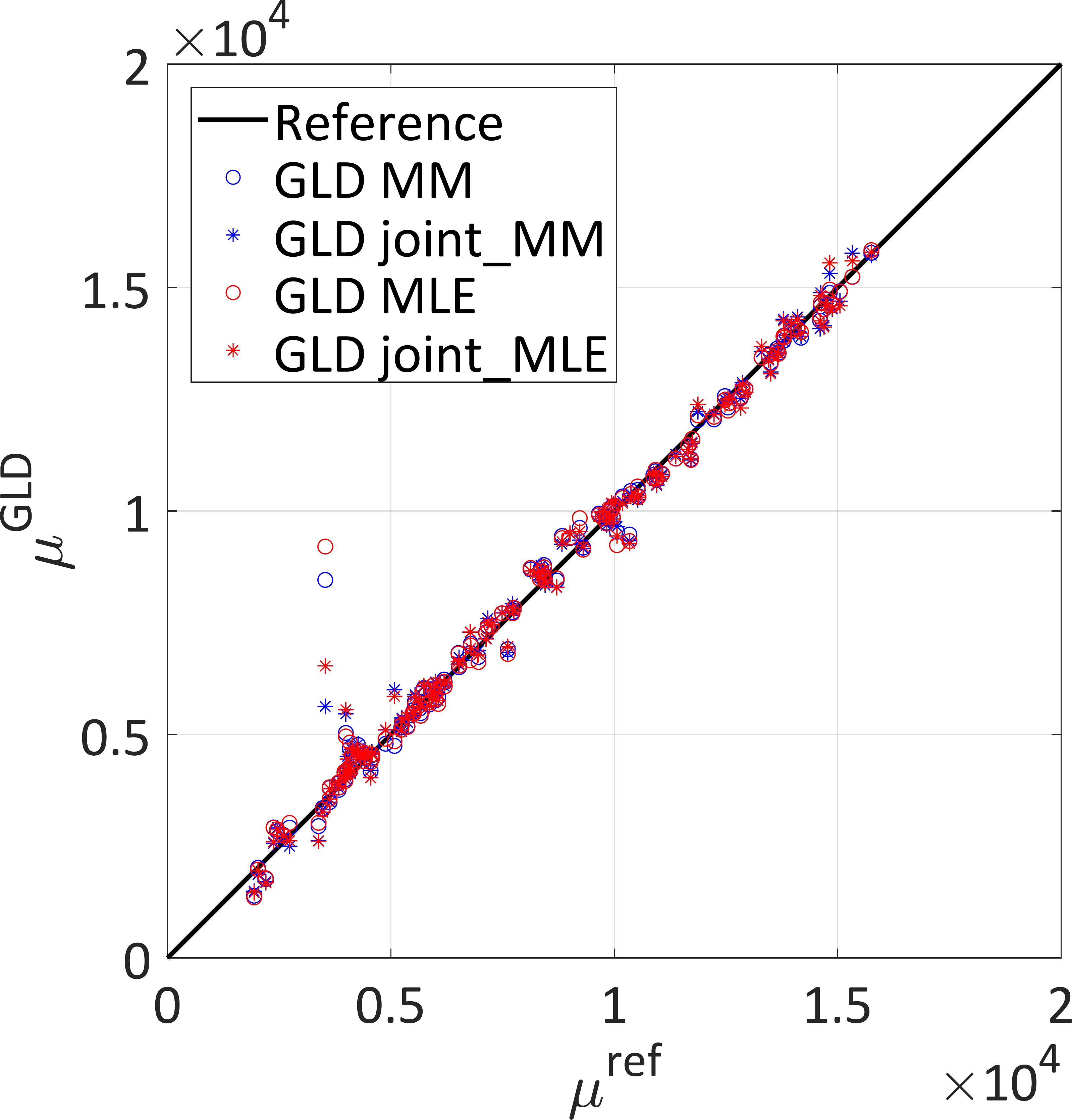}}
	\hspace{1cm}
	\subfigure[Std estimation]{\includegraphics[height=.4\linewidth, keepaspectratio]{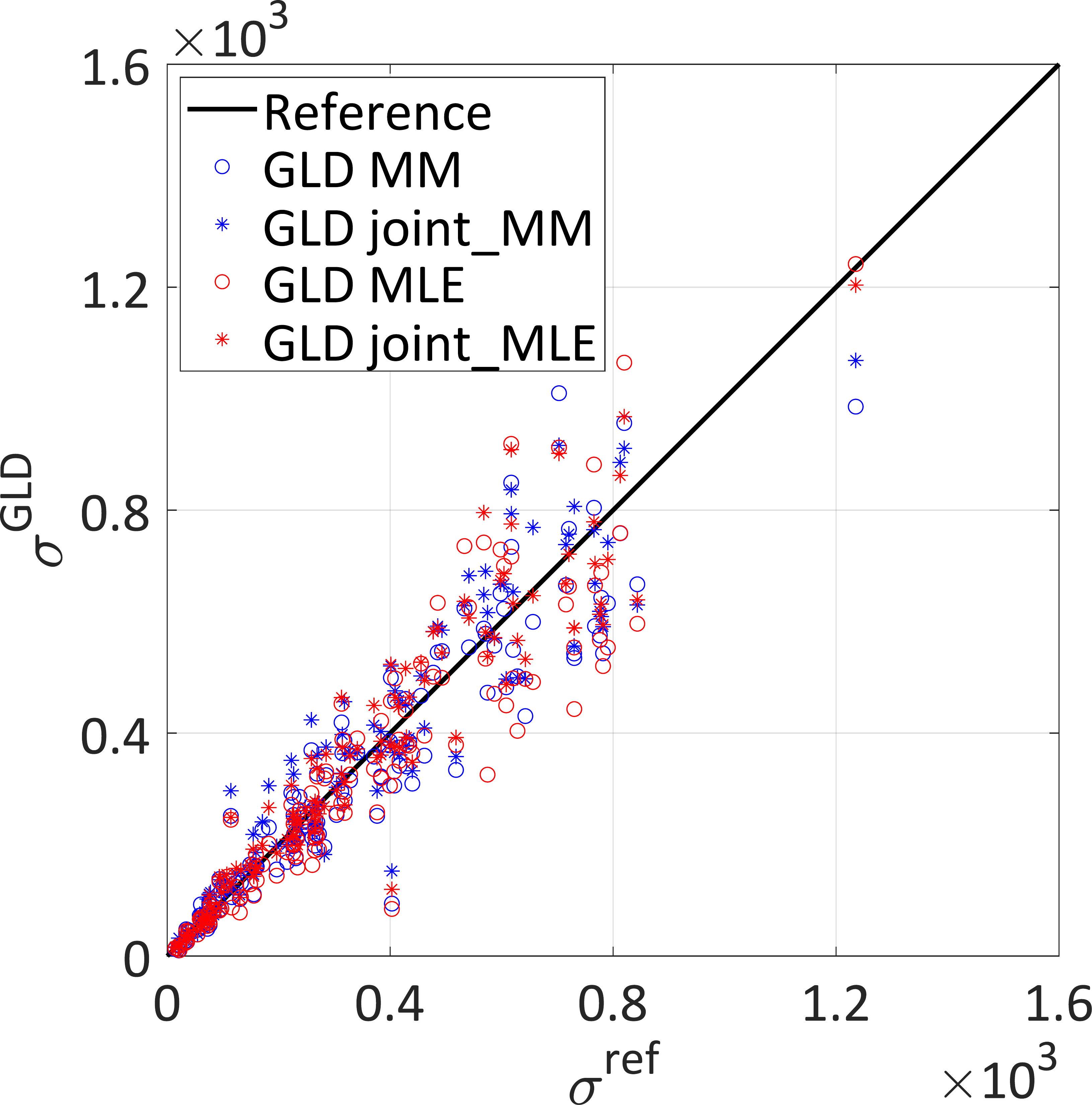}}
	\caption{Wind turbine case study -- Comparison of the mean and the standard deviation estimation of the maximum flapwise bending moment ($kN\cdot m$). The $x$-axis (reference) is the empirical quantity calculated from the 500 replications.}
	\label{fig:windms}
\end{figure}
\begin{figure}[!htbp]
	\centering
	\subfigure[5\% quantile estimation]{\includegraphics[height=.4\linewidth, keepaspectratio]{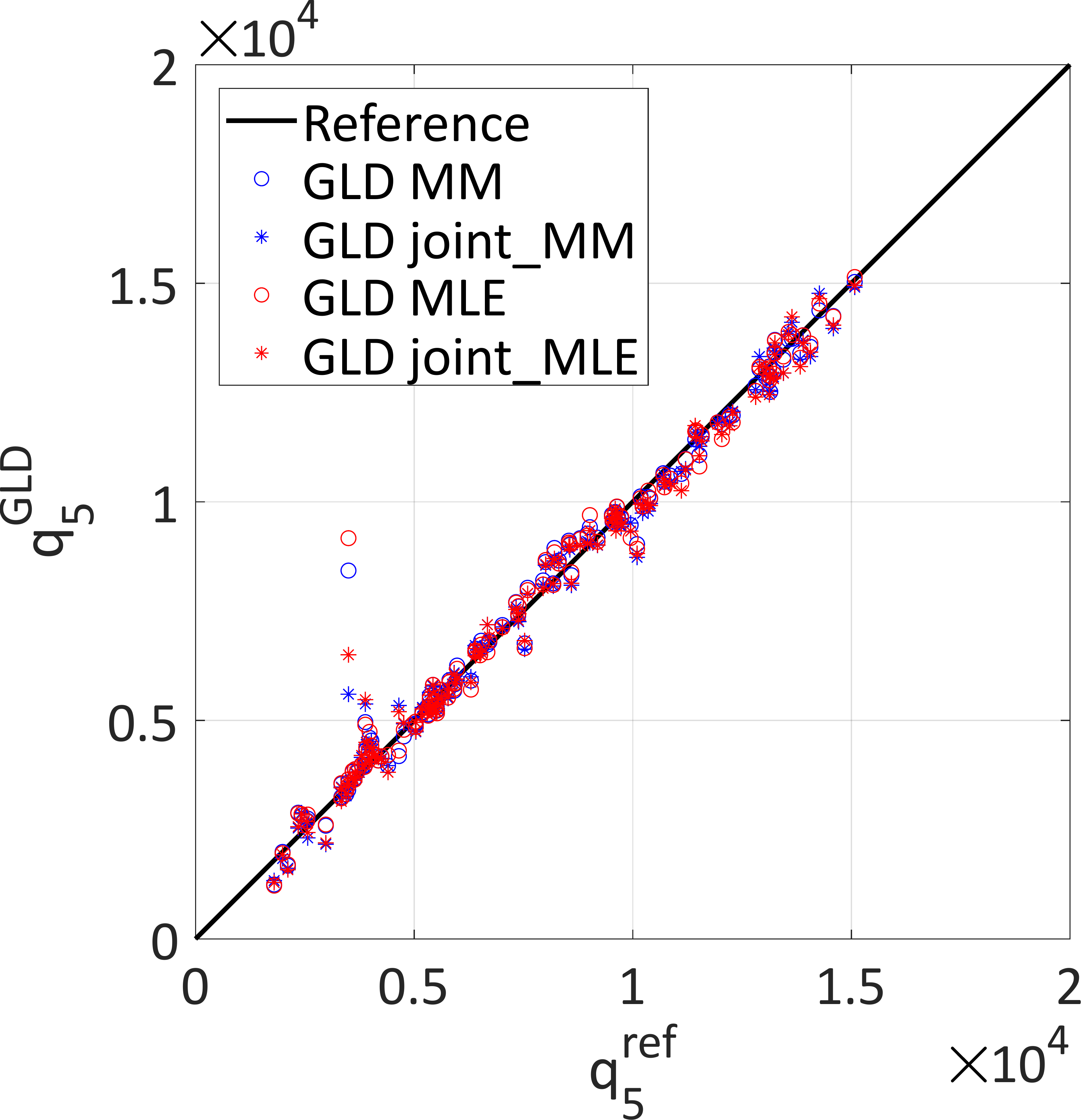}}
	\hspace{1cm}
	\subfigure[10\% quantile estimation]{\includegraphics[height=.4\linewidth, keepaspectratio]{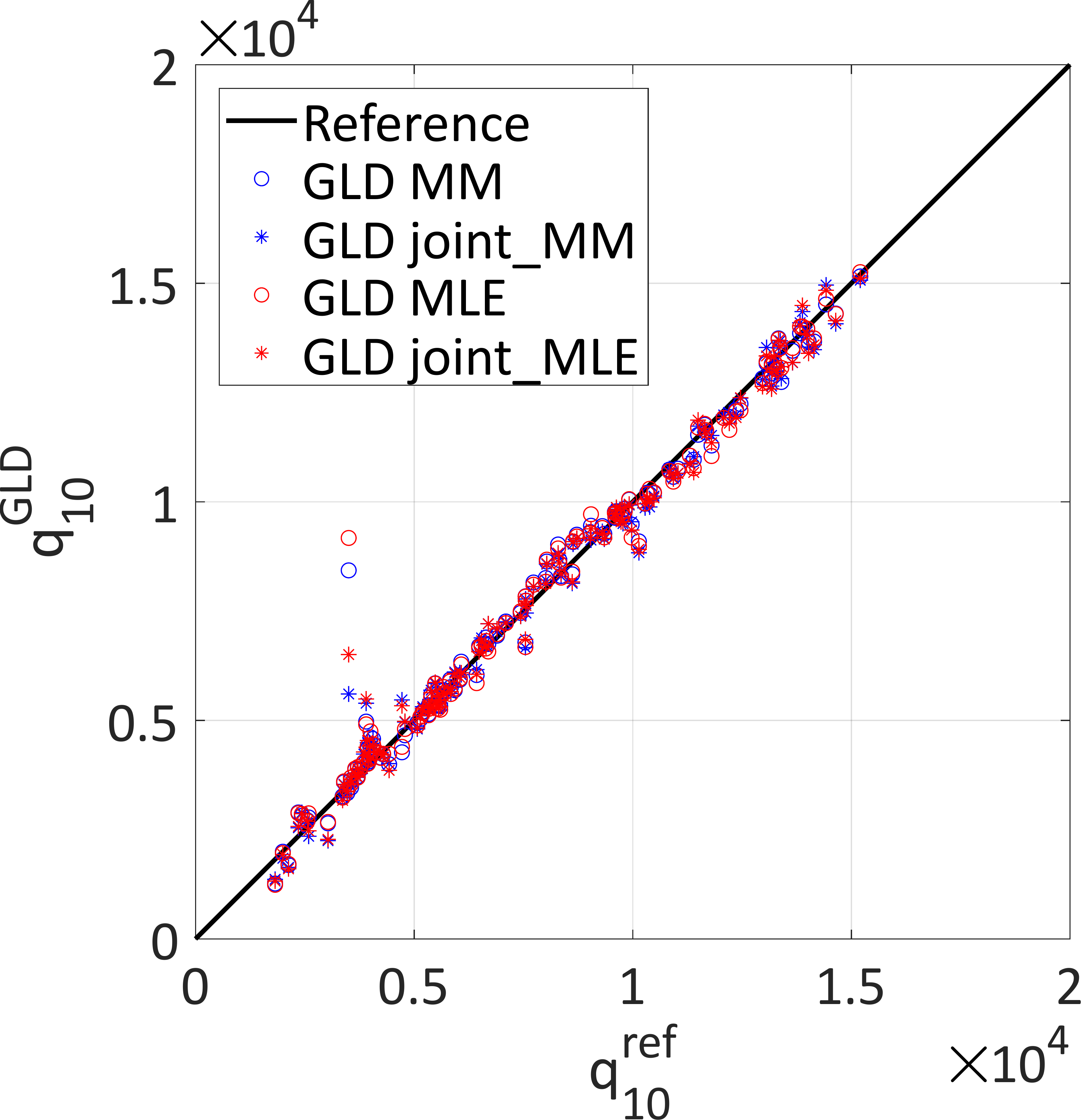}}
	\subfigure[50\% quantile estimation]{\includegraphics[height=.4\linewidth, keepaspectratio]{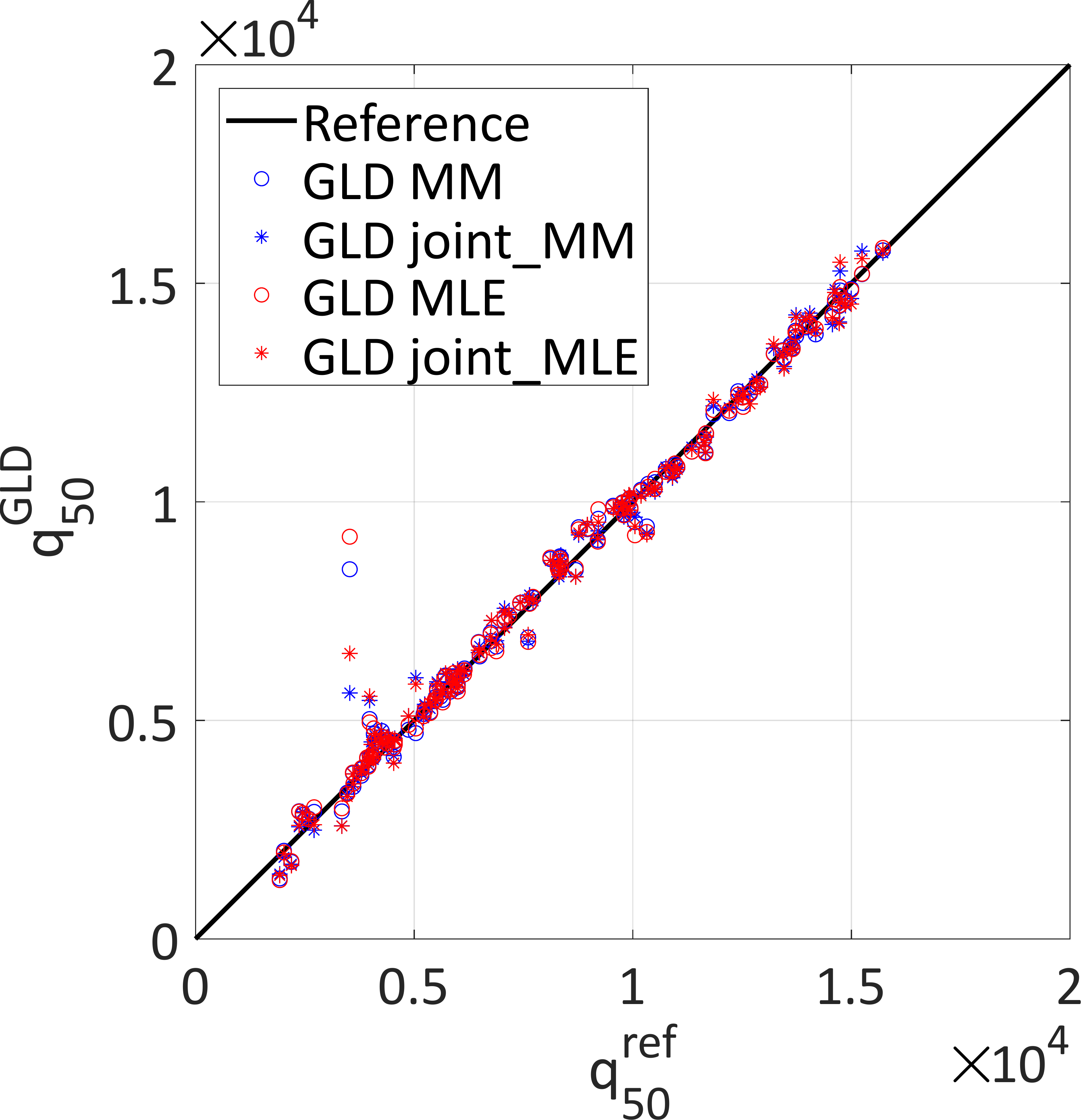}}
	\hspace{1cm}
	\subfigure[90\% quantile estimation]{\includegraphics[height=.4\linewidth, keepaspectratio]{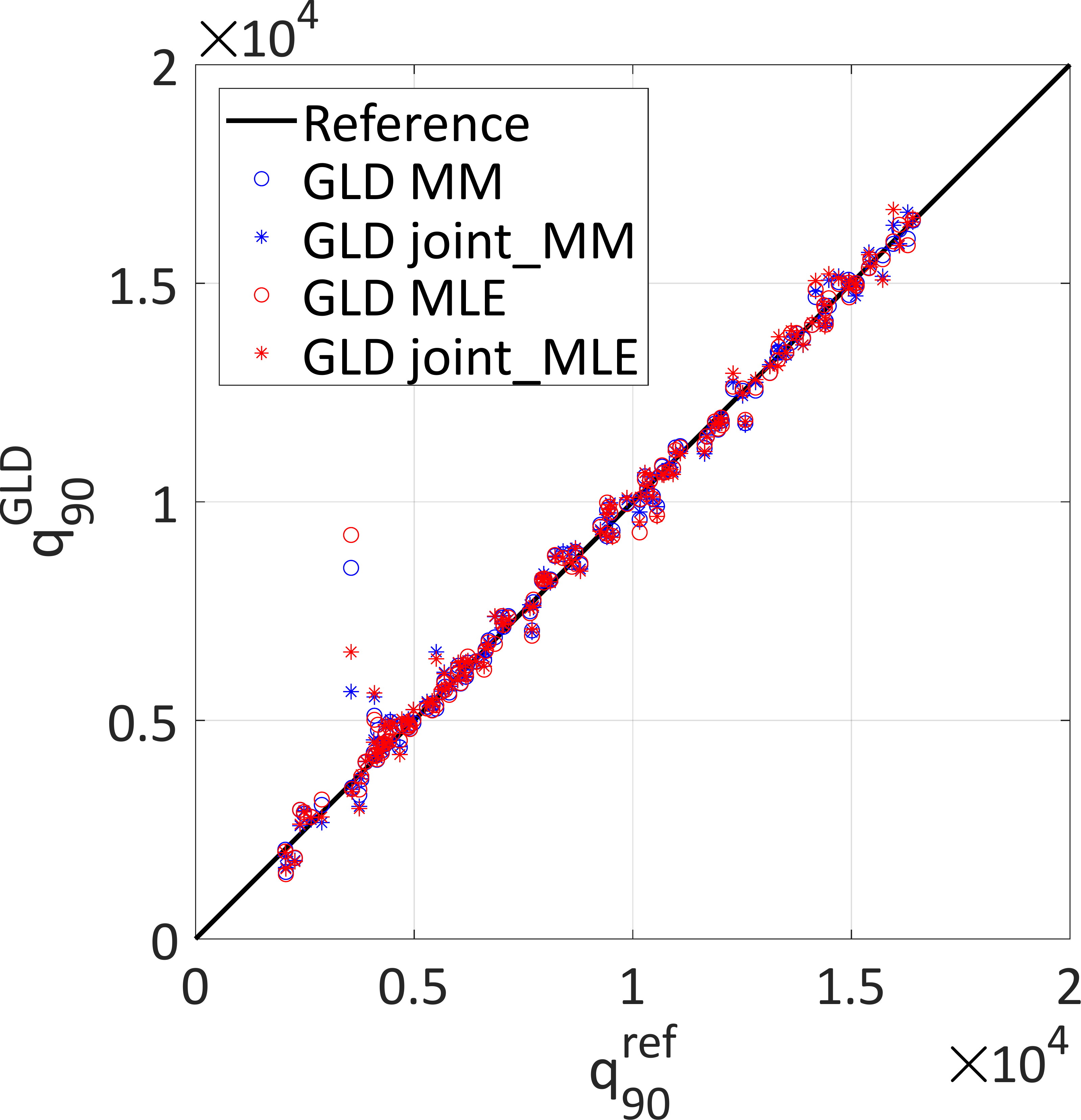}}
	\subfigure[95\% quantile estimation]{\includegraphics[height=.4\linewidth, keepaspectratio]{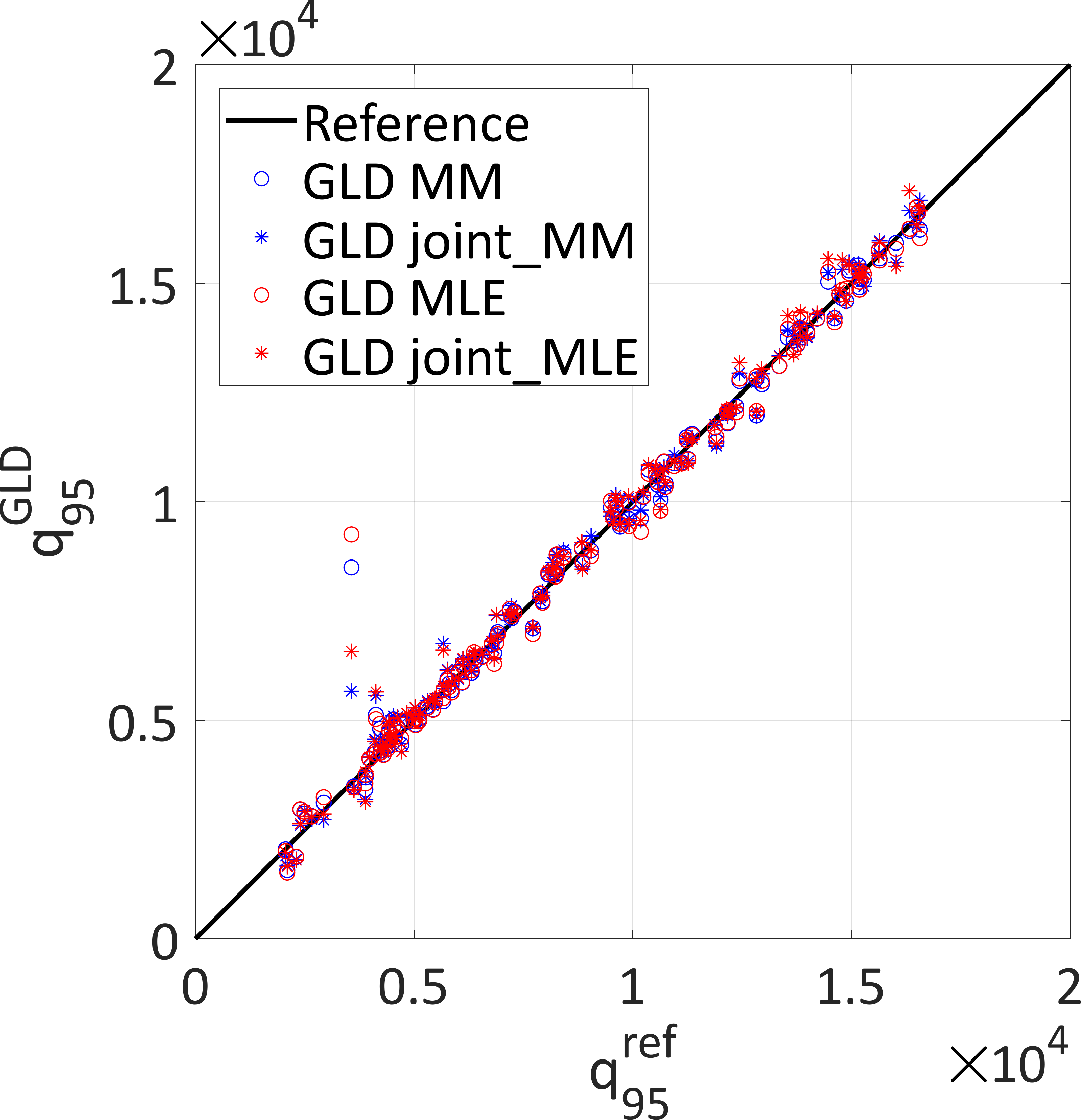}}
	\caption{Wind turbine case study -- Comparison of the quantiles estimation of the maximum flapwise bending moment ($kN\cdot m$). The $x$-axis (reference) is the empirical quantity calculated from the 500 replications.}
	\label{fig:windq}
\end{figure}

The results of the four GLD models are shown in \Cref{fig:windms} and \Cref{fig:windq}. Comparisons of the scalar quantities show that all the four GLD models demonstrate a good fit to the simulated scenario. Among the scalar quantities, the mean and the quantiles are estimated with high accuracy, whereas the standard deviation estimation is relatively poor.
\par
\par
To have more quantitative comparison among the four GLD models, we define the normalized mean squared error:
\begin{equation} 
\epsilon = \frac{\sum_{i=1}^{N_{\rm test}} \left(q^{(i)}_{GLD} - \hat{q}^{(i)}\right)^2}{\sum_{i=1}^{N_{\rm test}}\left(\hat{q}^{(i)} - \bar{\hat{q}}\right)^2 }, \text{ with }
\bar{\hat{q}} = \frac{1}{N_{\rm test}} \sum_{i=1}^{N_{\rm test}} \hat{q}^{(i)},
\end{equation}
where $q$ is the statistical quantity of interest (mentioned above), $q^{(i)}_{GLD}$ is the value predicted by the GLD model, and $\hat{q}^{(i)}$ denotes the estimated quantity (empirical mean, standard deviation and quantiles) based on the replications for $\ve{x}^{(i)}$. 
\par
The errors associated with the scalar quantities are reported in \Cref{tab:MSEq}. We can observe that the method of moments outperforms the maximum likelihood estimation for both the Infer-and-Fit and the joint model in terms of all the error measures used here. The joint models generally improve their associated Infer-and-Fit models, and \GJMM\ provides the best estimates.
\par
\begin{table}
	\centering
	\caption{Normalized mean squared error of various quantities in the test set. The best results among the four GLD models are highlighted in bold.}
	\label{tab:MSEq}
	\footnotesize
	\begin{tabular}{  c c c c c c c c}
		\hline
		GLD models & mean & std & $Q_{05}$ & $Q_{10}$ & $Q_{50}$ & $Q_{90}$ & $Q_{95}$\\
		\hline
		\GMM  & 0.0185  & \bestresult{0.1125} & 0.0231 & 0.0221 & 0.0188 & 0.0166 & 0.0165\\
		\GJMM  & \bestresult{0.0099}  & 0.124 & \bestresult{0.0133} & \bestresult{0.0154} & \bestresult{0.0103} & \bestresult{0.009} & \bestresult{0.0091}\\
		\GMLE  & 0.0235  & 0.1488 & 0.0292 & 0.0280 & 0.0237 & 0.0214 & 0.0213\\
		\GJMLE  & 0.0128  & 0.1642 & 0.0167 & 0.0155 & 0.0131 & 0.0121 & 0.0125\\
		\hline
	\end{tabular}
\end{table}
\par
\begin{figure}[!htbp]
	\centering
	\subfigure[PDF at $\ve{x} = \left(4.9,1.2,0.6,0.9,7.2\right)^T$]{\includegraphics[height=.35\linewidth, keepaspectratio]{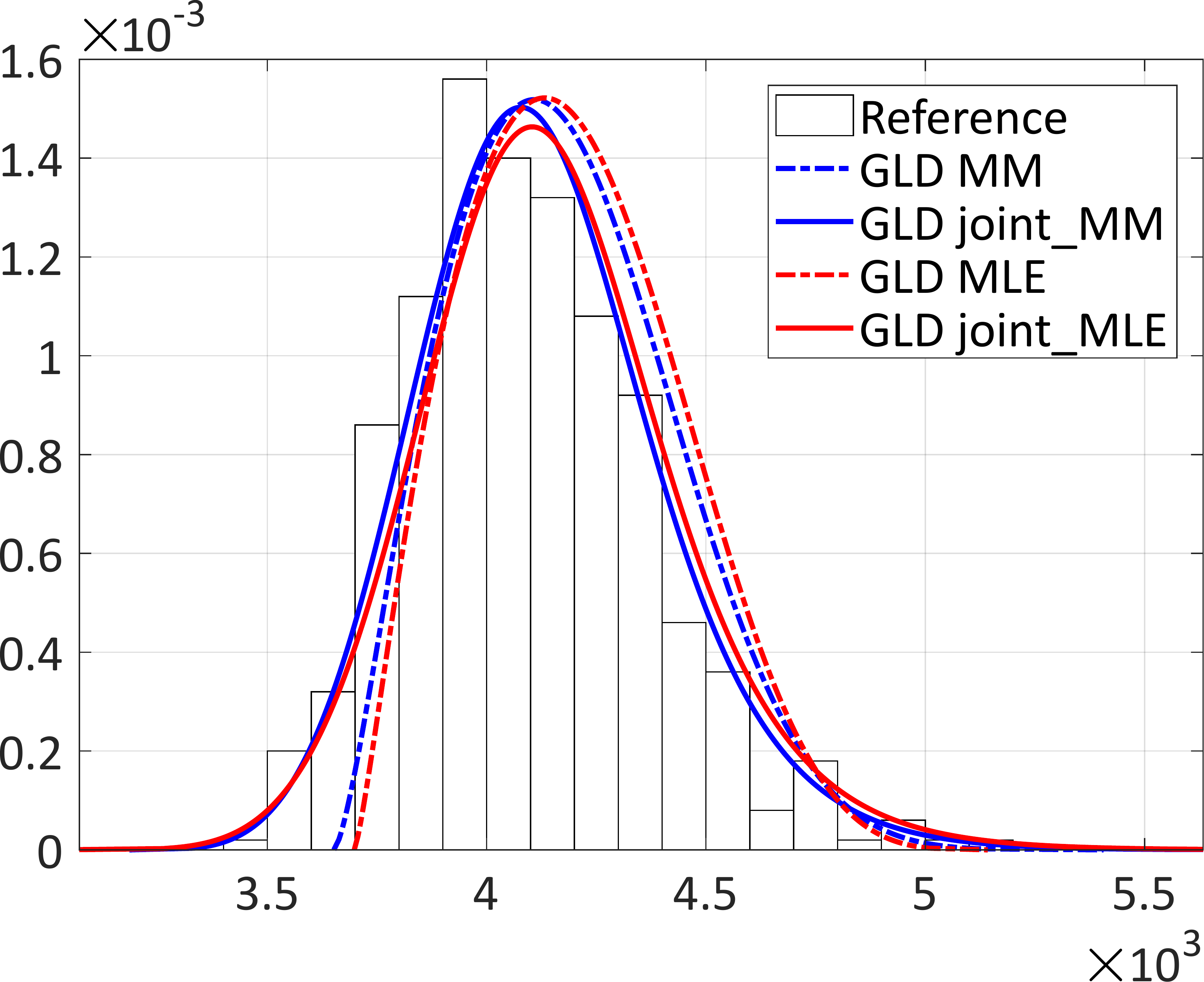}}
	\hspace{1cm}
	\subfigure[PDF at $\ve{x} = \left(11.8,2.9,0.3,1.0,0.2\right)^T$]{\includegraphics[height=.35\linewidth, keepaspectratio]{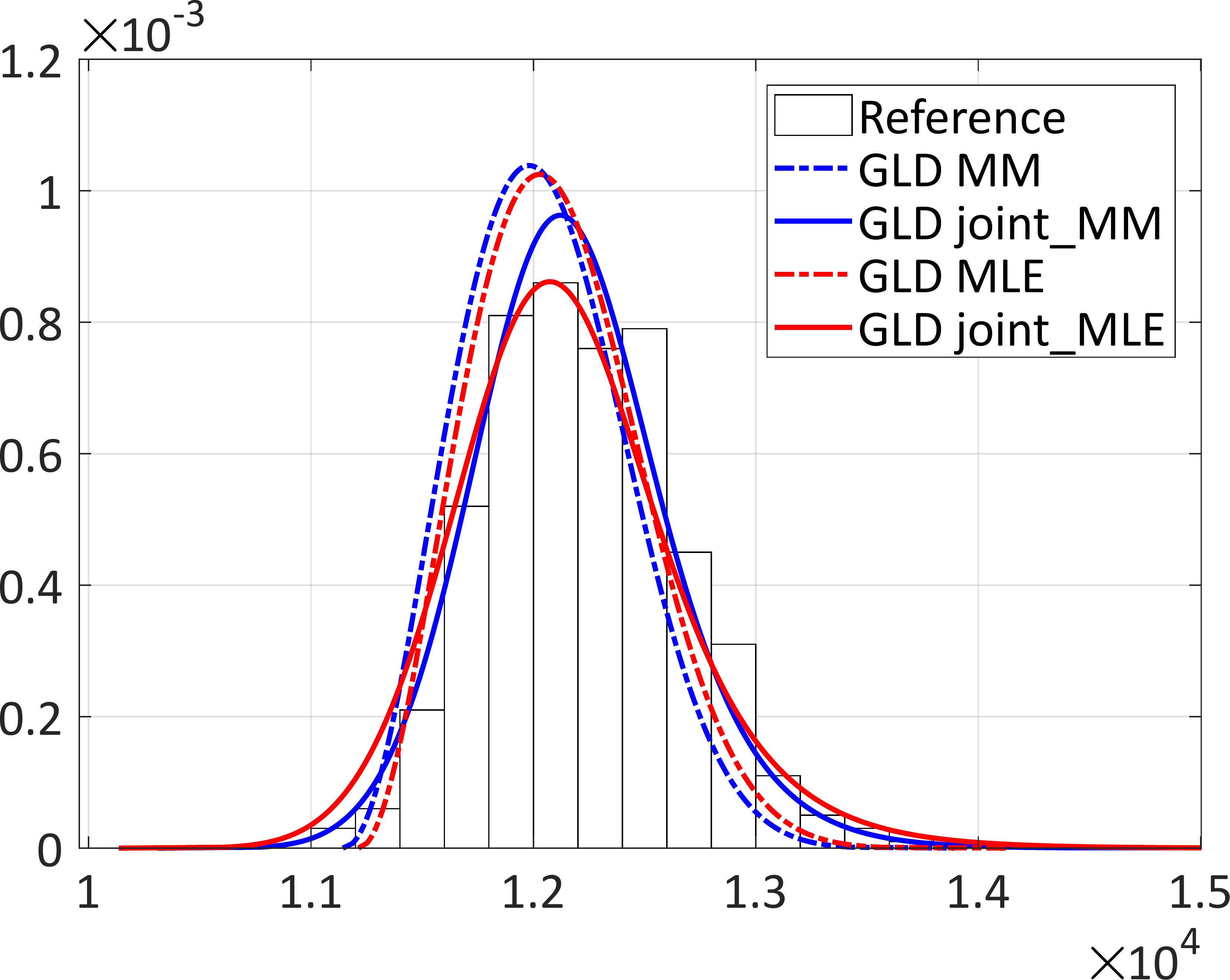}}
	\caption{Wind turbine case study -- PDF predictions with the experimental design of size 485 using 50 replications. The reference histogram is calculated based on 500 replications.}
	\label{fig:windpdf}
\end{figure}

Apart from the detailed quantitative comparison of the scalar quantities, we visualize also the PDF prediction in \Cref{fig:windpdf} for two specific values of $\ve{x}$. The reference histograms are obtained based on the 500 replications. Since only 500 samples are available, the histograms are less smooth than those of the previous example in \Cref{sec:sde}. We observe that all the four surrogate models can well capture the location of the underlying distribution. In addition, the two joint models demonstrate better performance on the shape approximation. For example, in \Cref{fig:windpdf}, the two Infer-and-Fit models produce narrower support than the range of the samples, whereas the support is accurately approximated by the two joint models. 
\par
As a conclusion, the GLD joint models allows for accurate prediction of the PDF of the maximum flapwise bending moment at the blade root at a total cost of about 24,000 runs of Tubsim+FAST. The calculation have been carried out on the ETH Euler cluster using 96 cores for a physical time of about 20.5 hours. Interestingly, the 95\% quantile, which is of interest for design assessment, is remarkably predicted all over the space of input parameters.

\section{CONCLUSIONS}
\label{sec:conclusions}
The aim of this paper is to build efficient and accurate surrogate models for stochastic simulators within the replication-based framework. Generalized lambda distributions are used to flexibly approximate the output PDF, while the relationship of their parameters with the inputs is approximated through polynomial chaos expansions. To construct surrogate models in a non-intrusive manner, we first proposed the Infer-and-Fit algorithm which consists of solving two consecutive problems. In the first step, the distribution parameters are inferred based on repeated model evaluations for each point of the experimental design. Then the estimated values are used to build a PCE surrogate model for each distribution parameters. The Infer-and-Fit algorithm allows us to use conventional regression techniques to construct PCE. However, this method is sensitive to the number of replications due to the two-step strategy, whereby the model responses are only used in the first step. In order to build accurate surrogate models even when a few replications are available, we proposed in a second part the joint modeling method described in \Cref{alg:joint}. This approach carries out one more optimization step after getting a first estimate from the Infer-and-Fit approach, which is used to provide sparse truncation sets for $\ve{\lambda}^{\PC}(\ve{x};\ve{a})$ and a starting point for the optimization. This enrichment allows us to use all the available data at once and provides a maximum likelihood estimator of the model parameters, namely the coefficients of the polynomial chaos expansions of the $\ve{\lambda}$'s. Due to the complexity of the likelihood function, this additional optimization problem can be expensive to solve. To alleviate the computational burden, we vectorized the implementation in the Matlab environment and derived analytically the gradient and the Hessian matrix of the objective function. As a result, we can efficiently apply derivative-based optimizers.
\par
For the analytical examples in \Cref{sec:examples} and the stochastic differential equation case study in \Cref{sec:sde}, the proposed two algorithms are both able to approximate the reference distributions with high accuracy, even though the data generation scheme does not follow the generalized lambda distribution. As expected, the joint models show better performance when only a few replications are available. For the wind turbine application in \Cref{sec:wind}, due to the cost of the simulator, only some important statistical scalar quantities are compared to a reference solution obtained from a large Monte Carlo simulation, whereas PDFs at selected input points are only compared visually. Both developed methods demonstrate high accuracy for the mean and quantiles estimation. 
\par
In all the examples and applications, joint models are observed to consistently improve the result of the associated models built with the Infer-and-Fit algorithm. Besides, for the parametric estimation in the first step of the Infer-and-Fit algorithm, the method of moments and maximum likelihood show comparable performance. This observation matches the conclusion in \cite{Corlu2016}.
\par
In the joint modeling method, the main role played by the replications is to obtain a truncation scheme for each component of $\ve{\lambda}^{\PC}(\ve{x};\ve{a})$ as well as to find an initial starting point for the following optimization step. Therefore, replications are not necessary if the basis functions of each distribution parameter are known or preselected. Work is in progress to improve the proposed method by using advanced statistical techniques that completely avoid the need for replications and thus drastically reduce the computational cost.

\section*{Acknowledgments}

This paper is a part of the project ``Surrogate Modeling for Stochastic Simulators (SAMOS)'' funded by the Swiss National Science Foundation (Grant \#200021\_175524). The authors gratefully thank Dr. Imad Abdallah (ETH Zurich) and Ren\'e Slot (Aalborg University) for extensive fruitful discussions on applications, and Nora L\"uthen (ETH Zurich) for the computation of the wind turbine data.

\bibliography{References}
\appendix
\section{Appendix}

\subsection{Feasible starting point}
\label{sec:feasible}
For the additional optimization problem introduced in the joint algorithm, the coefficients $\tilde{\ve{a}}$ fitted from the Infer-and-Fit algorithm are chosen as an appropriate starting point for the optimization. However, as discussed in \Cref{sec:JM}, the objective function $l(\ve{a})$ can take the value $+\infty$, and thus complex constraints are present, which is summarized in \refEq{eq:constraints}. Therefore, additional operations are necessary to have a feasible starting point if $\tilde{\ve{a}}$ does not satisfy the constraints. 

It is observed from \refEq{eq:Bounds} that the lower (upper) bound of the support is a monotonic function of $\lambda_3$ ($\lambda_4$) for fixed $\lambda_1$ and $\lambda_2$. Therefore, reducing the coefficients $a_{3,\ve{0}}$ and $a_{4,\ve{0}}$ that are associated with the constant functions in \refEq{eq:lamPCE_2} broadens the support of the response PDF for all $\ve{x} \in \cd_{\ve{X}}$. Based on this property, the following procedure is proposed to adjust $\tilde{\ve{a}}$ to be feasible:
\begin{enumerate}
	\item Evaluate $\ve{\lambda}^{(i)} = \ve{\lambda}^{\PC}\left(\ve{x}^{(i)};\tilde{\ve{a}}\right)$ for all $\ve{x}^{(i)} \in \ve{\cx}$
	\item Collect the index $i$ into the set $I$, whose associated $\lambda^{(i)}_3$ is positive 
	\item For all $i \in I$, calculate $\check{\lambda}^{(i)}_3$ such that the minimum value of the associated replication results located exactly on the lower bound. More precisely, according to \refEq{eq:Bounds}, we have
	\begin{equation}
	\check{\lambda}^{(i)}_3 = \frac{1}{\lambda^{(i)}_2\left(\lambda^{(i)}_1 - \min\limits_{r} y^{(i,r)}\right)}
	\end{equation}
	\item Decrease the value of $\tilde{a}_{3,\ve{0}}$ so that $\lambda^{(i)}_3 < \check{\lambda}^{(i)}_3$ for all $i \in I$. 
\end{enumerate}
This algorithm only deals with the constraints related to the lower bounds. The same method can be used for those from upper bounds, which consists in modifying the constant $\tilde{a}_{4,\ve{0}}$ based on $\tilde{\lambda}_4^{(i)}$.

\subsection{Analytical derivations for \Cref{alg:joint}}
\label{sec:Deriv}
In this section, we compute the analytical derivatives of the negative log-likelihood function $l$ with respect to the model parameters $\ve{a}$. Since the objective function is a composition of several functions as shown in \Cref{fig:network}, the derivatives can be calculated through the chain rule, which flows from Step 3 to Step 1. Starting from
\begin{equation}
l=\log\left(\frac{u^{\lambda_3-1}+(1-u)^{\lambda_4-1}}{\lambda_2}\right),
\end{equation}
we get the following partial derivatives:
\begin{align}
\frac{\partial l}{\partial u} &= \frac{ (\lambda_3-1)u^{\lambda_3-2} - (\lambda_4-1)(1-u)^{\lambda_4-2}}{u^{\lambda_3-1} +(1-u)^{\lambda_4-1}}, \label{eq:dldu}\\
\frac{\partial l}{\partial \lambda_2} &= -\frac{1}{\lambda_2}, \label{eq:dldlam2}\\
\frac{\partial l}{\partial \lambda_3} &= \frac{u^{\lambda_3-1}\log(u)}{u^{\lambda_3-1} +(1-u)^{\lambda_4-1}}, \label{eq:dldlam3}\\
\frac{\partial l}{\partial \lambda_4} &= \frac{ (1-u)^{\lambda_4-1}\log(1-u)}{u^{\lambda_3-1} + (1-u)^{\lambda_4-1}}. \label{eq:dldlam4}
\end{align}

The differentiation at Step 2 is more complex because $u$ is not an explicit function of $\ve{\lambda}$, and thus it involves derivatives of a highly nonlinear implicit function \refEq{eq:FKML}. Based on
\begin{align*}
y = Q(u) = \lambda_1 + \frac{1}{\lambda_2}\left( \frac{u^{\lambda_3}-1}{\lambda_3} - \frac{(1-u)^{\lambda_4}-1}{\lambda_4}\right),
\end{align*}
and because $y$ is given, differentiating both side gives
\begin{equation*}
0 = \D \left(\lambda_1 + \frac{1}{\lambda_2}\left( \frac{u^{\lambda_3}-1}{\lambda_3} - \frac{(1-u)^{\lambda_4}-1}{\lambda_4}\right)\right),
\end{equation*}
where $\D$ stands for the total differentiation.

Expanding and rearranging the equations above, we have
\begin{align}
\frac{\partial u}{\partial \lambda_1} = & -\frac{\lambda_2}{u^{\lambda_3-1} + (1-u)^{\lambda_4 - 1}}, \label{eq:dudlam1}\\
\frac{\partial u}{\partial \lambda_2} = & \frac{1}{\lambda_2 \left(u^{\lambda_3-1} + (1-u)^{\lambda_4 - 1}\right)} \left( \frac{u^{\lambda_3}-1}{\lambda_3} - \frac{(1-u)^{\lambda_4}-1}{\lambda_4}\right), \label{eq:dudlam2}\\
\frac{\partial u}{\partial \lambda_3} = & \frac{u^{\lambda_3}-\lambda_3u^{\lambda_3}\log(u)-1}{\lambda^2_3\left(u^{\lambda_3-1} + (1-u)^{\lambda_4 - 1}\right)}, \label{eq:dudlam3}\\
\frac{\partial u}{\partial \lambda_4} = & \frac{-(1-u)^{\lambda_4}+\lambda_4(1-u)^{\lambda_4}\log(1-u)+1}{\lambda^2_4\left(u^{\lambda_3-1} + (1-u)^{\lambda_4 - 1}\right)}.  \label{eq:dudlam4}
\end{align}
As illustrated in \Cref{fig:network}, the derivatives of the negative log-likelihood function with respect to $\ve{\lambda}$ come from two parts: one is from the direct derivative (\refEqs{eq:dldlam2}{eq:dldlam4}) in Step 3, the other part is contributed by the implicit differentiation (\refEqs{eq:dudlam1}{eq:dudlam4}) in Step 2. As a result, we have
\begin{align}\label{eq:dldlam}
\frac{\D l}{\D \lambda_s} &= \frac{\partial l}{\partial \lambda_s} + \frac{ \partial l}{\partial u}\frac{\partial u}{\partial\lambda_s} , \quad s=1,2,3,4.
\end{align}
Finally, the derivatives flow back to the model parameters $\ve{a}$ at Step 1 as follows:
\begin{align}
\frac{\D l}{\D a_{s,\ve{\alpha}}} = \frac{\D l}{\D \lambda_s}\frac{\partial \lambda_s}{\partial a_{s,\ve{\alpha}}} &= \frac{\D l}{\D \lambda_s} \psi_{\ve{\alpha}}\left(\ve{x}\right) , \quad s=1,3,4, \label{eq:dlhdlam134}\\
\frac{\D l}{\D a_{2,\ve{\alpha}}}=\frac{\D l}{\D \lambda_2}\frac{\partial \lambda_2}{\partial a_{2,\ve{\alpha}}} &=\frac{\D l}{\D \lambda_2}\frac{1}{L^{\prime}\left(\lambda_2\right)} \psi_{\ve{\alpha}}\left(\ve{x}\right), \label{eq:dlhdlam2}
\end{align}
where $L$ denotes the transform that is used to guarantee the positiveness of $\lambda_2(\ve{x})$. Recall that we chose to use $L(\lambda_2) = \log(\lambda_2)$ in this paper. Similar techniques can be used to derive the Hessian matrix of the negative log-likelihood function, which is necessary for the trust-region algorithm. Due to the lengthy derivation, we omit the result here. \refEq{eq:dldlam} calculates the derivatives of the log-likelihood function with respect to the distribution parameters $\ve{\lambda}$. Hence, it can be used in the maximum likelihood estimation of the distribution parameters of a random variable following a generalized lambda distribution.

\end{document}